\documentclass[manuscript]{acmart}
\usepackage{booktabs}
\usepackage{threeparttable}
\usepackage{multirow}
\usepackage[inline]{enumitem}
\usepackage{subcaption}
\usepackage{amsthm}
\usepackage{enumitem}
\usepackage{tikz}
\usepackage{amsfonts}
\usepackage{algorithm}
\usepackage{bm}
\usepackage[noend]{algpseudocode}
\usetikzlibrary{bayesnet}
\usepackage{xcolor}
\usepackage[utf8]{inputenc}

\newcommand{\bb}[1]{\bm{#1}}

\newcommand{\br}{\bb{r}}
\newcommand{\bhr}{\bb{\hat{r}}}
\newcommand{\bR}{\bb{R}}

\newcommand{\bhRzero}{\bb{\hat{R}}^0}
\newcommand{\bhRone}{\bb{\hat{R}}^1}
\newcommand{\bru}{\br_{u}}
\newcommand{\bhru}{\bhr_{u}}
\newcommand{\bhrut}{\bhr_{u}^{t}}

\newcommand{\bhruzero}{\bhr_{u}^{0}}
\newcommand{\brui}{\br_{u,i}}

\newcommand{\bk}{\bb{k}}
\newcommand{\bhk}{\bb{\hat{k}}}
\newcommand{\bK}{\bb{K}}
\newcommand{\bKI}{\bb{K^I}}
\newcommand{\bku}{\bk_{u}}

\newcommand{\bkIi}{\bk^{I}_{i}}
\newcommand{\bkIic}{\bk^{I}_{i,\bc}}
\newcommand{\bhku}{\bhk_{u}}

\newcommand{\bz}{\bb{z}}
\newcommand{\btz}{\bb{\tilde{z}}}

\newcommand{\bzu}{\bz_{u}}

\newcommand{\bzuc}{\bz_{u}^{c}}
\newcommand{\bzuct}{\bz_{u}^{t}}
\newcommand{\btzuct}{\btz_{u}^{t}}

\newcommand{\bc}{\bb{c}}

\newcommand{\bcu}{\bc_{u}}
\newcommand{\bcut}{\bc_{u}^{t}}

\newcommand{\bx}{\bb{x}}
\newcommand{\bmu}{\bb{\mu}}
\newcommand{\bS}{\bb{\Sigma}}
\newcommand{\bmuu}{\bmu_{u}}
\newcommand{\bSu}{\bS_{u}}

\copyrightyear{2021}
\acmYear{2021}
\setcopyright{acmlicensed}\acmConference[RecSys '21]{Fifteenth ACM Conference on Recommender Systems}{September 27-October 1, 2021}{Amsterdam, Netherlands}
\acmBooktitle{Fifteenth ACM Conference on Recommender Systems (RecSys '21), September 27-October 1, 2021, Amsterdam, Netherlands}
\acmPrice{15.00}
\acmDOI{10.1145/3460231.3474249}
\acmISBN{978-1-4503-8458-2/21/09}

\author{Diego Antognini}
\email{firstname.lastname@epfl.ch}
\author{Boi Faltings}
\affiliation{%
  \institution{
École Polytechnique Fédérale de Lausanne}
  \city{Lausanne}
  \country{Switzerland}
}

\begin{document}

\title{Fast Multi-Step Critiquing for VAE-based Recommender~Systems}

\begin{abstract}

Recent studies have shown that providing personalized explanations alongside recommendations increases trust and perceived quality. Furthermore, it gives users an opportunity to refine the recommendations by \textit{critiquing} parts of the explanations.
On one hand,~current recommender systems model the recommendation, explanation, and critiquing objectives jointly, but this creates an inherent trade-off between their respective performance. 
On the other hand, although recent latent linear critiquing approaches are built~upon an existing recommender system, they suffer from computational inefficiency at inference due to the objective optimized at each~conversation's turn.
We address these deficiencies with M\&Ms-VAE, a novel variational autoencoder for recommendation and explanation that is based on multimodal modeling assumptions. We train the model under a weak supervision scheme to simulate both~fully~and partially observed variables. Then, we leverage the generalization ability of a trained M\&Ms-VAE model to embed the user preference and the critique separately. Our work's most important innovation is our critiquing module, which is built upon and trained in~a~self-supervised manner with a simple ranking objective.
Experiments on four real-world datasets demonstrate that among state-of-the-art models, our system is the first to dominate or match the performance in terms of recommendation, explanation, and multi-step critiquing. Moreover, M\&Ms-VAE processes the critiques up to 25.6x faster than the best baselines. Finally, we show that our model~infers coherent joint and cross generation, even under weak supervision, thanks to our multimodal-based modeling and training~scheme.

\end{abstract}


\maketitle

\section{Introduction}

Recommender systems accurately capture user preferences and achieve high performance. However, they offer little transparency regarding their inner workings. It has been shown that providing explanations along with item recommendations enables users to understand why a particular item has been suggested and hence to make better decision~\cite{chang2016crowd,bellini2018knowledge}. Additionally, explanations increase the system's overall transparency and trustworthiness~\cite{tintarev2015explaining,zhang2018exploring,kunkel2018trust}.

An important advantage of explanations is that they provide a basis for feedback. If users understand what has generated the suggestions, they can refine the recommendations by interacting directly with the explanations. Critiquing~is a conversational recommendation method that incrementally adapts recommendations in response to user preferences~\cite{chen2012critiquing}. Example critiquing was introduced in information retrieval~\cite{williams1982rabbit} and first applied to recommender systems in~\cite{unitcritiquing}. Recognizing that critiquing is most useful when applied in multiple steps,~\cite{Linden97interactiveassessment} and~\cite{10.1145/332040.332446} introduced mechanisms based on constraint programming~\cite{Torrens2002abcd} with an application to travel planning. Multi-step critiquing with constraint programming was recognized as a form of preference elicitation, which enabled the analysis and optimization of its performance~\cite{Faltings2004b} and the addition of suggestions for active preference elicitation~\cite{Viappiani06preference-basedsearch}, which yielded dramatic improvements in decision accuracy in user studies. Multi-step critiquing was also shown to be superior to compound critiquing, which groups multiple attributes in a single step~\cite{10.1145/1250910.1250929}. A major limitation of all these approaches is that items have to be characterized by a set of discrete attributes.

After nearly a decade in which critiquing approaches received little attention,~\cite{wu2019} introduced a collaborative filtering recommender with explanations and an embedding-based critiquing method. This method allows users to critique the recommendation using arbitrary languages; a set of attributes is mined from reviews, and the users can interact~with them. Other works built upon the same paradigm~\cite{antognini2020interacting,chen2020towards}.~\cite{luo2020} showed that those models suffer from unstable training and high computational complexity, and they proposed a framework based on a variational autoencoder~\cite{kingmamw2014,liang2018variational}. However, these models learn a bidirectional mapping between the critique and the user latent space. This creates~an inherent trade-off between the recommendation and explanation performance, and it yields poor results in~multi-step~critiquing.
 
Recently,~\cite{luo2020b} proposed a latent linear critiquing (LLC) method built upon the  recommendation model PLRec~\cite{sedhain2016practical}. LLC co-embeds keyphrase attributes in the same embedding space as the recommender. The critiquing process consists~of~a weighted average between the user-preference embedding and the critique embeddings obtained through the conversation. The weights are optimized in a linear programming formulation using a max-margin scoring-based objective~(i.e., the pairwise difference of scores of items affected by the critique and the others). Following the same methodology,~\cite{hanze2020} changed the objective into a ranking-based one. While those models obtain good performance in multi-step critiquing, they suffer from computational inefficiency due to the objective function optimized at each turn.

To address both issues, we present M\&Ms-VAE, a novel variational autoencoder for recommendation and explanation with a separate critiquing module. Inspired by multimodal generative models~\cite{NEURIPS2018_1102a326,NEURIPS2019_0ae775a8,tsai2018learning,NEURIPS2020_43bb733c}, we treat the user's past interactions and keyphrase usage as different partially observed variables, and more importantly, we assume conditional independence between them. We can then approximate the variational joint posterior using a mixture of experts. We propose a training scheme that mimics weakly supervised learning to train the inference networks jointly but also independently. This is essential to our modeling, because M\&Ms-VAE is robust to a missing unobserved variable and can thus embed separately and efficiently the user interactions, the keyphrases, and the critique,~respectively. 

In a second step, we leverage the generalization ability of M\&Ms-VAE and design a novel blending module to re-rank recommended items according to a critique. The latter is trained only once on a synthetic dataset with a self-supervision objective. This generalizes into multi-step critiquing and enables fast critiquing.

To the best of our knowledge, this is the first work to revisit deep critiquing from the perspective of multimodal~generative models and to propose a blending module trained in a simple self-supervised fashion. 
We evaluate our method~using four real-world datasets. The results demonstrate that the proposed M\&Ms-VAE model \begin{enumerate*}
 	\item achieves superior~or~competitive performance in terms of recommendation, explanation, and multi-step critiquing in comparison to the state-of- the-art recommendation and critiquing methods, 
 	\item processes the critiques up to 26x faster than the best baselines and up to 9x faster using only the CPU, and 
 	\item induces coherent joint and cross generation, even under weak supervision.
 \end{enumerate*}
 
\section{Preliminaries}
This section introduces the notation used in the paper and the variational autoencoder for recommendation~\cite{liang2018variational}. Then, we review a recent study~\cite{luo2020} that built upon~\cite{liang2018variational} and revisited critiquing by proposing the critiquable-explainable VAE (CE-VAE) model. Finally, we highlight the key deficiencies that significantly limit its performance in practice.
\subsection{Notation}
Before proceeding, we define the following notation used throughout this paper:
\begin{itemize}
	\item $U$, $I$, and $K$: The user, the item, and the keyphrase sets, respectively.
	\item $\bR \in \mathbb{R}^{|U|\times|I|}$: The user-by-item interaction matrix obtained with implicit feedback. Entries $\brui$ of~1~(respectively~0) denote a positive (respectively negative or unobserved) interaction between the user $u$ and item~$i$.
	\item $\bK \in \mathbb{R}^{|U|\times|K|}$: The binary user-keyphrase matrix that reflects the user $u$'s keyphrase-usage preference. Given user reviews from a corpus, we extract keyphrases that describe item attributes from all reviews (see Section~\ref{sec_datasets}).
	\item $\bKI \in \mathbb{R}^{|I|\times|K|}$: The binary item-keyphrase matrix. The process is similar to $\bK$ with the aggregation per item.
	\item $\bhru \in \mathbb{R}^{|I|}$ and $\bhku \in \mathbb{R}^{|K|}$: The predicted feedback and keyphrase explanation, respectively.
	\item $\bzu \in \mathbb{R}^{|H|}$: The user $u$'s latent embedding of dimension $H$ from the observed interaction $\bru$ and keyphrase-usage preference $\bku$.
	\item $\bcut \in \mathbb{R}^{|K|}$: A one-hot vector of length $|K|$. The only positive value indicates the index of the keyphrase to be critiqued by the user $u$ at a given step $t$ of the user interaction with the recommender system.
	\item $\bzuct \in \mathbb{R}^{|H|}$: The latent representation of the critique $\bcut$.
	\item $\btzuct \in \mathbb{R}^{|H|}$: The updated latent representation of the user after the critique $\bcut$.
	\item $I^{+\bc} \in \{i|\bkIic=1, \forall i \in I\}$: The set of items that contain the critiqued keyphrase $\bc$.
	\item $I^{-\bc} \in \{i|\bkIic=0, \forall i \in I\}$: The set of items that do not contain the critiqued keyphrase $\bc$.
\end{itemize}

\subsection{Variational Autoencoder for Recommendation (VAE)}
\label{sec_vae}
A variational autoencoder (VAE)~\cite{kingmamw2014} is a generative model of the form $p_\theta(\bx, \bz) = p(\bz)p_\theta(\bx | \bz)$, where $p(\bz)$ is a prior and the likelihood $p_\theta(\bx | \bz)$ is parametrized by a neural network with parameters $\theta$. The model learns to maximize the marginal likelihood of the data $p_\theta(\bx)$ (i.e., the evidence that is intractable) by approximating the true unknown posterior $p_\theta(\bz | \bx)$ with a variational posterior $q_\phi(\bz | \bx)$. Applied to recommendation systems, the collaborative-filtering VAE~\cite{liang2018variational} considers as input data the sparse user preferences $\bru$ over $|I|$ items. More formally, the model optimizes a variational lower bound on the log likelihood of all observed user feedback $\sum_{u \in U} \log p(\bru)$ through stochastic gradient~descent: \begin{equation}
	 \log p(\bru) \ge \int_{\bzu} q_\phi(\bzu | \bru) \log \frac{p_\theta(\bru, \bzu)}{q_\phi(\bzu | \bru)} d\bzu \ge \mathbb{E}_{q_\phi(\bzu | \bru)} \bigl[\log p_\theta(\bru | \bzu)\bigr] - \beta \textrm{ D}_{\textrm{KL}} \bigl[ q_\phi(\bzu | \bru) ~||~ p(\bzu) \bigr],
\end{equation}
where $\bzu$ is sampled\footnote{Using the reparametrization trick~\cite{kingmamw2014,10.5555/3044805.3045035}: $\bzu = \bmuu + \epsilon \bb{\sigma}_u$, where $\epsilon \sim \mathcal{N}(0, \mathbb{I}_H)$.} from the distribution $q_\phi(\bzu | \bru)$ with parameters $\bmuu$ and $\bSu$, and $\textrm{D}_{\textrm{KL}}[q, p]$ denotes the Kullback-Leibler divergence (KL) between the distributions $p$ and $q$. In practice, the prior $p(\bz)$ is usually a spherical Gaussian with parameters $\bmu$ and $\bS$. Finally, $\beta$ is a hyperparameter that controls the strength of the regularization relative to the reconstruction error, as motivated by the $\beta$-VAE of~\cite{higgins2016beta}, and is slowly annealed to 1, similarly to~\cite{bowman-etal-2016-generating}.

\subsection{Co-embedding of Language-based Feedback with the Variational Autoencoder (CE-VAE)} \label{sec_cevae}
Thus far, the variational autoencoder can only recommend items without generating any form of explanation. A recent study~\cite{luo2020} proposed the CE-VAE model, which integrates an explanation and critiquing module based on keyphrases. The authors support critiquing by first modeling the joint probability of a user's item preferences and keyphrase usage: \begin{align}
\log p(\bru, \bku) = \log p(\bku | \bru) + \log p(\bru) &= \mathbb{E}_{q_{\Phi_r}(\bzu | \bru)} \bigl[\log p_{\Theta_k}(\bku | \bzu)\bigr] - \textrm{D}_{\textrm{KL}} \bigl[ q_{\Phi_r}(\bzu | \bru) ~||~ p(\bzu) \bigr]\\
&+  \mathbb{E}_{q_{\Psi_r}(\bzu | \bru)} \bigl[\log p_{\Theta_r}(\bru | \bzu)\bigr] + \mathcal{H}\bigl[q_{\Psi_r}(\bzu | \bru)\bigr] + \mathbb{E}_{q_{\Psi_r}(\bzu | \bru)} \bigl[p(\bzu) \bigr], \nonumber
\end{align}
where $\mathcal{H}$ is the entropy. Then, they incorporate an additional objective to learn a projection from the critiquing feedback into the latent space via another encoder (an inverse feedback loop). In other words, they reintroduce the user's keyphrase usage $\bku$ to approximate the variational lower bound of $p(\bzu)$ by marginalizing over $\bku$. More formally:\begin{align}
\log p(\bzu) &\ge \mathbb{E}_{q(\bku | \bzu)} \bigl[\log p(\bzu | \bku)\bigr] - \textrm{D}_{\textrm{KL}} \bigl[ q(\bku | \bzu) ~||~ p(\bku) \bigr]\\
&\approx \mathbb{E}_{p_{\Theta_k}(\bku | \bzu)} \bigl[\log p_{\Theta_k'}(\bzu | \bku)\bigr] - \textrm{D}_{\textrm{KL}} \bigl[ p_{\Theta_k}(\bku | \bzu) ~||~ p(\bku) \bigr],\nonumber
\end{align}
where $p(\bku)$ is a prior following a standard normal distribution and the weights of $q(\bku | \bzu)$ are shared with $p_{\Theta_k}(\bku | \bzu)$.

Finally, once the model is trained on the full objective function, the critiquing process for the critique $\bcu$ is performed as follows: 
\begin{enumerate*}
	\item compute the critique representation $\bzuc$ with $p_{\Theta_k'}(\bzu | \bku)$,
	\item average both the user latent representation~$\bzu$ and the critique representation $\bzuc$, and 
	\item predict the new feedback $\bhru$ with the generative network~$p_{\Theta_r}(\bru | \bzu)$.
\end{enumerate*}

Overall, the CE-VAE framework is effective in practice for recommendation, keyphrase explanation, and single-step critiquing. However, it suffers from two key deficiencies that limit its performance (as we later show empirically):\begin{enumerate}
	\item The model learns a function to project the critiqued keyphrase into the user's latent space, from which the~feedback and the explanation are predicted. This mapping is learned via an autoencoder, which perturbs the training. Thus, there is an inherent trade-off between the performance of the recommendation and that of the~explanation.
	\item Although the joint objective also maximizes a latent representation likelihood with the Kullback-Leibler terms, it is unclear~whether the inverse function embeds the critique effectively and whether the mean reflects a critiquing mechanism.
\end{enumerate}

\section{M\&Ms-VAE: A Mixture-of-Experts Multimodal Variational AutoEncoder}
Our goal is to build a more generalizable representation of users' preferences that is based on their observed interactions and keyphrase usage. 
Figure~\ref{fig_gm} depicts the graphical model of our proposed M\&Ms-VAE, and Figure~\ref{fig_training} shows the training scheme. 
Then, we leverage this representation to efficiently embed the user critiques and learn, in a self-supervised fashion, a blending module to re-rank recommended items for multi-step critiquing. Figure~\ref{fig_critiquing_general} illustrates the workflow.

\subsection{Model Overview}
Like previously developed variational autoencoders for recommendation, we assume that the observed~user $u$'s interactions~$\bru$ and the keyphrase-usage preference $\bku$ are generated from a latent representation of the user preferences.

Differently from prior work, we seek to learn the joint distribution $p(\bru,\bku)$ under weak supervision. Our main~goal~is to learn a more generalizable representation of the user preferences. Therefore, we aim to design a generative model~that can recommend and generate keyphrase explanation \textbf{jointly but also independently} from each of the observed~variables (i.e., cross-modal generation). It also allows us to apply the same technique to users who have not written~reviews or to cases in which keyphrases are unavailable.
 If this goal is achieved, we can then embed effectively the~user's observed interactions, the user's keyphrase preference, and the critique with the \textbf{same} inference network~$q_\Phi(\bzu | \bru, \bku)$.

Inspired by multimodal generative models~\cite{NEURIPS2018_1102a326,NEURIPS2019_0ae775a8,tsai2018learning,NEURIPS2020_43bb733c}, we treat $\bru$ and $\bku$ as different modalities, and we assume they are conditionally independent given the common latent variable $\bzu$. In other words, we assume a generative model of the form $p_\Theta(\bru, \bku, \bzu)=p(\bzu)p_{\Theta_r}(\bru|\bzu)p_{\Theta_k}(\bku|\bzu)$. An advantage of such a factorization is that if $\bru$ or~$\bku$ is unobserved, we can safely ignore it when evaluating the marginal likelihood~\cite{NEURIPS2018_1102a326}.

\begin{figure*}[]
\centering
\begin{minipage}[t]{0.45\textwidth}
  \centering
  \begin{tikzpicture}
  	  \node[latent] (z) {$\bz$};
      \node[latent, below=of z, xshift=-0.75cm,path picture={\fill[gray!25] (path picture bounding box.south) rectangle (path picture bounding box.north west);}] (r) {$\br$} ;
      \node[latent, below=of z, xshift=0.75cm,path picture={\fill[gray!25] (path picture bounding box.south) rectangle (path picture bounding box.north west);}] (k) {$\bk$} ;
      
	  \node[xshift=1.9cm] (1) {$\Phi_k$};
	  \node[xshift=-1.9cm] (2) {$\Phi_r$};
	  \node[below=of z, yshift=-0.1cm, xshift=1.9cm] (3) {$\Theta_k$};
	  \node[below=of z, yshift=-0.1cm, xshift=-1.9cm] (4) {$\Theta_r$};
    \plate[inner sep=0.3cm, xshift=0cm, yshift=0cm] {plate} {(r) (k) (z)} {$u \in \{1 \dots U\}$};
	\path[->] (z) edge [bend left=20] node {} (r) ;
	\path[->] (z) edge [bend left=-20] node {} (k) ;
	\path[->, dashed] (r) edge [bend left=20] node {} (z) ;
	\path[->, dashed] (k) edge [bend left=-20] node {} (z) ;
	
	\path[->, dashed] (1) edge [] node {} (z) ;
	\path[->, dashed] (2) edge [] node {} (z) ;
	\path[->] (3) edge [] node {} (k) ;
	\path[->] (4) edge [] node {} (r) ;
  \end{tikzpicture}
    \caption{\label{fig_gm}Probabilistic-graphical-model view of our proposed M\&Ms-VAE model. Both the implicit feedback $\bru$ and the keyphrase $\bku$ are generated from user $u$'s latent representation $\bzu$. Solid lines denote the generative model, whereas dashed lines denote the variational approximation.}
\end{minipage}
\hfill
\begin{minipage}[t]{0.52\textwidth}
    \centering
    \includegraphics[width=1.0\textwidth,height=1.37in]{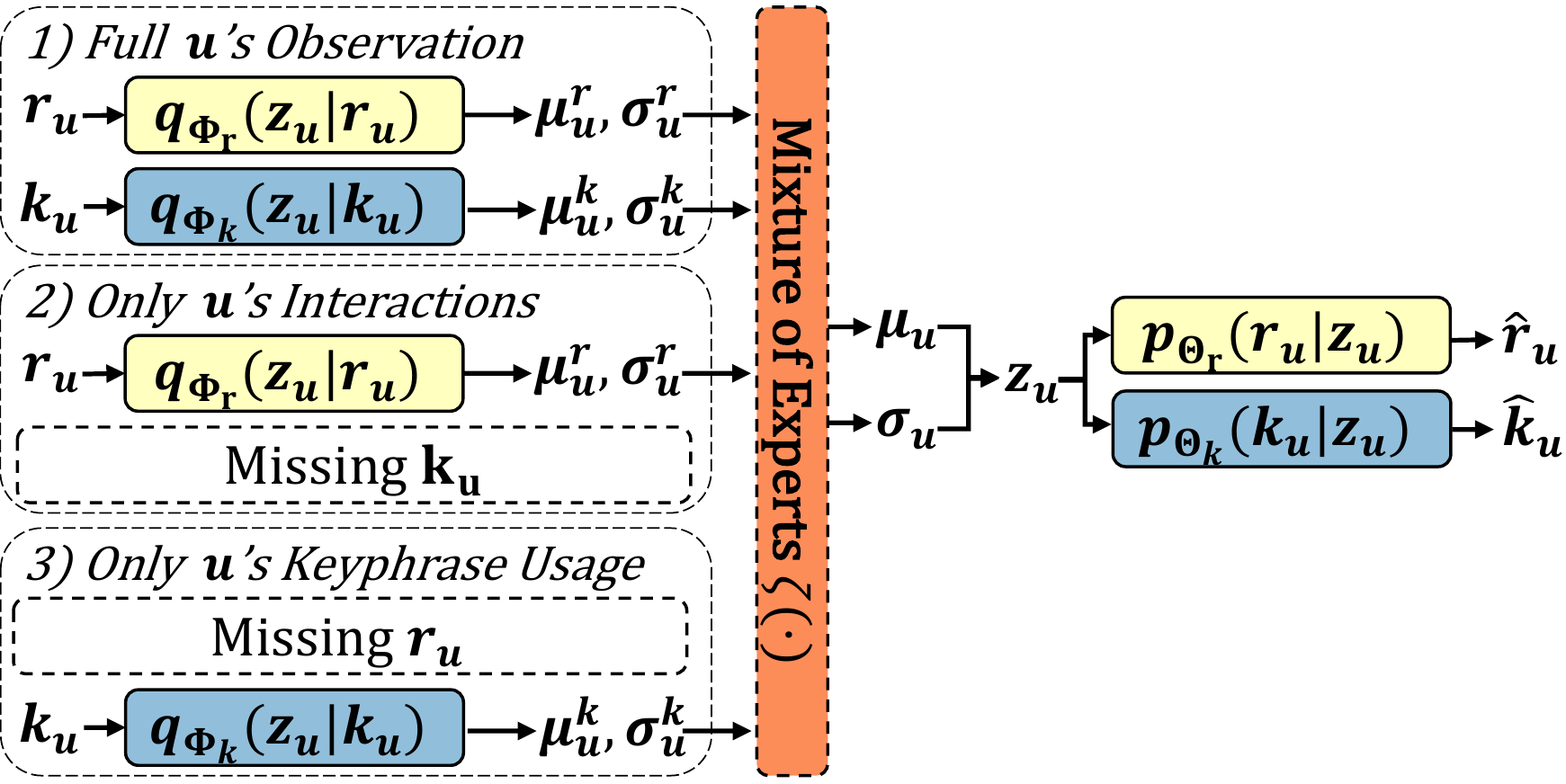}
    \caption{\label{fig_training}The proposed M\&Ms-VAE architecture and training scheme. Each pass infers the parameters $\bmuu$ and $\bb{\sigma}_u$ with the mixture of experts using either the joint inference network $q_\Phi(\bzu | \bru, \bku)$ or one of the individual networks ($q_{\Phi_r}(\bzu | \bru)$ or $q_{\Phi_k}(\bzu | \bku)$, respectively). The final gradient is computed on the sum of each $ELBO(\cdot)$ term.}
\end{minipage}
\end{figure*}

We start with the derivation of the joint log likelihood $\sum_{u \in U} \log p(\bru, \bku)$ over the observed interactions~$\bru$ and keyphrase-usage preference $\bku$ and all users $u$ as shown in Figure~\ref{fig_gm}:\begin{equation}
\label{eq_mm_1}
\log~p(\bru, \bku) = \log \int_{\bzu} p_\Theta(\bru, \bku, \bzu) d\bzu \ge \mathbb{E}_{q_\Phi(\bzu | \bru, \bku)} \bigl[\log p_\Theta(\bru, \bku | \bzu)\bigr] - \beta \textrm{ D}_{\textrm{KL}} \bigl[ q_\Phi(\bzu | \bru, \bku) ~||~ p(\bzu) \bigr],
\end{equation} where we assume that the prior distribution $p(\bz)$ is a standard normal distribution and $\beta$ is a hyperparameter that controls the strength of the regularization relative to the reconstruction error.
Thanks to our assumption that $\bru$ and~$\bku$ are conditionally independent given the common latent variable $\bzu$, we can rewrite Equation~\ref{eq_mm_1} as follows:
\begin{equation}
	\label{eq_mm_2}
	ELBO(\bru, \bku) = \mathbb{E}_{q_\Phi(\bzu | \bru, \bku)} \bigl[\log p_{\Theta_r}(\bru | \bzu) + \log p_{\Theta_k}(\bku | \bzu)\bigr] - \beta \textrm{ D}_{\textrm{KL}} \bigl[ q_\Phi(\bzu | \bru, \bku) ~||~ p(\bzu) \bigr].
\end{equation}

Learning the variational joint posterior $q_\Phi(\bzu | \bru, \bku)$ of Equation~\ref{eq_mm_2} under its current form requires $\bru$ and $\bku$ to be presented at all times, thus making cross-modal recommendation difficult. Following our assumption, we can factorize the joint variational posterior as a function $\zeta(\cdot)$ of unimodal posteriors (or experts) $q_{\Phi_r}(\bzu | \bru)$ and $q_{\Phi_k}(\bzu | \bku)$: $q_\Phi(\bzu | \bru, \bku) = \zeta\bigl(q_{\Phi_r}(\bzu | \bru), q_{\Phi_k}(\bzu | \bku)\bigr)$, similarly to~\cite{NEURIPS2018_1102a326,NEURIPS2019_0ae775a8,tsai2018learning,NEURIPS2020_43bb733c}. In our case, the function $\zeta(\cdot)$ should be \begin{enumerate*}
 \item robust to overconfident experts if the marginal posterior $q_{\Phi_r}(\bzu | \bru)$ or $q_{\Phi_k}(\bzu | \bku)$ has low density, and
 \item robust to missing unobserved variable $\bru$ or $\bku$.
\end{enumerate*} Therefore, we propose to rely on a mixture of experts (MoE) with uniform weights: \begin{equation}
\label{eq_moe}
	\zeta\bigl(q_{\Phi_r}(\bzu | \bru), q_{\Phi_k}(\bzu | \bku)\bigr) = \alpha \cdot q_{\Phi_r}(\bzu | \bru) + (1 - \alpha) \cdot q_{\Phi_k}(\bzu | \bku) \textrm{ with } \alpha = \begin{cases}
    \frac{1}{2}, & \text{if } \bru \textrm{ and } \bku \textrm{ are observed,}\\
    1, & \text{if only } \bru \textrm{ is observed,}\\
    0, & \text{if only } \bku \textrm{ is observed.}\\
\end{cases}
\end{equation}
We set the weights uniformly to explicitly enforce an equal contribution from each $\bru$ and $\bku$ when both are observed during training. In the case of an unobserved modality, we shift the importance distribution toward the presented one, which generalizes to weakly supervised learning (see Section~\ref{seq_training}). This is an important factor, because the inference network $q_{\Phi_k}(\bzu | \bku)$ will later induce the critique representation. Finally, one might be tempted to learn $\alpha$ jointly with the variational lower bound or dynamically. However, doing so might miscalibrate the precisions of the $q_{\Phi_r}(\bzu | \bru)$ or $q_{\Phi_k}(\bzu | \bku)$ and thus be detrimental to the whole model in terms of both prediction performance and generalization.

\subsection{Training Strategy}
\label{seq_training}
Combining Equations~\ref{eq_mm_2} and \ref{eq_moe} gives the full objective function, and M\&Ms-VAE can be trained on a complete dataset where all $\bru$ and $\bku$ are provided. However, in doing so, we never train the individual inference networks $q_{\Phi_r}(\bzu | \bru)$ and $q_{\Phi_k}(\bzu | \bku)$; only the relationship between the observed user interactions and keyphrase-usage preferences is captured. As a consequence, at inference, it is unclear how the model performs with a missing observation.
\\
To reach our goal of recommending given at least $\bru$ and embedding the critique effectively with the inference network $q_{\Phi_k}(\bzu | \bku)$, we propose a training strategy that mimics weakly supervised learning, similarly to~\cite{NEURIPS2018_1102a326}. Moreover, this allows us to handle incomplete datasets, where some samples are partially observed: data that contain only $\bru$~or $\bku$.\footnote{It also enables another way to solve the cold-start problem: new users can select a set of items and/or relevant keyphrases that reflect their preferences.} The training strategy is shown in Figure~\ref{fig_training}. For each minibatch, we compute the gradient on the evidence lower bound of the joint observation and each single observation $\bru$ and $\bku$. Our final training objective for all users $u$ is {\small \begin{align}
	\label{eq_mm_final}
	\mathcal{L}(\bR, \bK) &= \sum_{u \in U}\lambda \cdot \mathbb{E}_{q_\Phi(\bzu | \bru, \bku)} \bigl[\log p_{\Theta_r}(\bru | \bzu) + \log p_{\Theta_k}(\bku | \bzu)\bigr] - \beta \textrm{ D}_{\textrm{KL}} \bigl[ q_\Phi(\bzu | \bru, \bku) ~||~ p(\bzu) \bigr]  && \tag*{$ELBO(\bru,\bku)$} \nonumber \\
	&+ \sum_{u \in U} \lambda \cdot \mathbb{E}_{q_{\Phi_r}(\bzu | \bru)} \bigl[\log p_{\Theta_r}(\bru | \bzu)\bigr] - \beta  \textrm{ D}_{\textrm{KL}} \bigl[ q_{\Phi_r}(\bzu | \bru) ~||~ p(\bzu) \bigr] && \tag*{$ELBO(\bru)$} \nonumber \\
	&+ \sum_{u \in U} \lambda \cdot \mathbb{E}_{q_{\Phi_k}(\bzu | \bku)} \bigl[\log p_{\Theta_k}(\bku | \bzu)\bigr] - \beta \textrm{ D}_{\textrm{KL}} \bigl[ q_{\Phi_k}(\bzu | \bku) ~||~ p(\bzu) \bigr], && \text{$ELBO(\bku)$} 
\end{align}}where $\lambda$ and $\beta$ control the strength of the reconstruction error and regularization, respectively.

\subsection{Self-Supervised Critiquing with M\&Ms-VAE}

\begin{figure*}
  \centering
  \begin{minipage}[t]{0.5\textwidth}
  \centering
  \includegraphics[width=\textwidth]{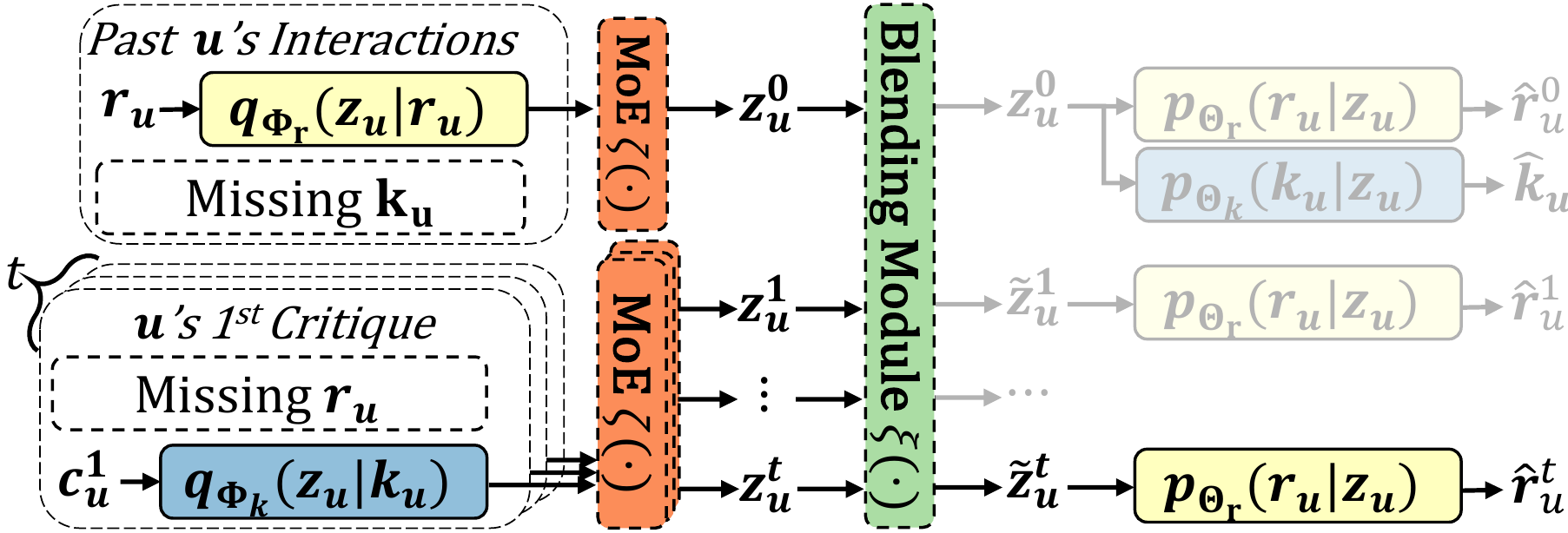}
  \caption{\label{fig_critiquing_general}Workflow of considering the recommendation of items~to~a user $u$ over $t$+1 time steps. First, M\&Ms-VAE produces the~initial set of recommended items $\bhruzero=\bhru$ using only the historical observed interactions $\bru$. Then, the user can provide a critique~$\bcut$ that is encoded into $\bzuct$ via the inference model~$q_{\Phi_k}(\bzu | \bku)$.~The blending module combines the previous representations~$\bz_{u}^{0}$,~$\bz_{u}^{1}$, $\dots, \bzuct$ into $\btzuct$, from which the subsequent recommendation~$\bhrut$ is computed. This process continues until the user $u$~accepts the recommendation and ceases to provide additional critiques.}
\end{minipage}
\hfill
\begin{minipage}{0.46\textwidth}
    \centering
\begin{algorithm}[H]
\caption{\label{alg_dataset}Synthetic Critiquing Dataset Creation}
\begin{algorithmic}[1]
\Function{Generate}{$\bR^{\textrm{val}}, \bKI$}
	\State Synthetic dataset $D \leftarrow \{\}$
    \For{each user $u$}
    	\For{each target item $i$, where $\brui^{\textrm{val}}=1$}
        \State Randomly sample a critique $\bc \in K \backslash\bkIi$
        \State Compute the item sets $I^{+\bc}$ and $I^{-\bc}$
        \State Update $D \leftarrow D \cup \{(u, i, \bc, I^{+\bc}, I^{-\bc})\}$
        \EndFor
    \EndFor
    \State \textbf{return} Synthetic dataset $D$
\EndFunction
\end{algorithmic}
\end{algorithm}
\end{minipage}
\end{figure*}
The purpose of critiquing is to refine the recommendation $\bhru$ based on the user $u$'s interaction with the explanation $\bhku$, represented with a binary vector. The user can accept the recommended items, at which point the session terminates. In the other case, the user can provide a critique $\bc_{u}^{t}$ and obtain a new recommendation $\bhrut$. The process is repeated over $T$ iterations until the user $u$ is satisfied with the recommendation. Each critique $\bcut$ is encoded as a one-hot vector where the positive value indicates a keyphrase the user $u$ dislikes. The overall process is depicted in Figure~\ref{fig_critiquing_general}.

We leverage the generalization ability of the trained M\&Ms-VAE, especially the inference models $q_{\Phi_r}(\bzu | \bru)$ and $q_{\Phi_k}(\bzu | \bku)$. We use the former to represent the initial user preferences $\bru$ and the latter to embed the critique~$\bcut$. However, a crucial question remains: how should we blend the user representation $\bzu$ with the $t$\textsuperscript{th} critique representation ~$\bzuct$?

Prior work has implemented a blending function as a simple average~\cite{wu2019,luo2020} or as a linear programming task that looks for a convex combination of embeddings provided with a specific linear optimization objective~\cite{luo2020b,hanze2020}. As we demonstrate empirically later, the former yields poor performance when iterated for multi-step critiquing, whereas the latter is computationally slow because the optimization is performed for each critique and it cannot leverage GPUs.

\subsubsection{Blending Function Design}
We propose to learn a blending function $\xi(\cdot)$ built upon a trained M\&Ms-VAE model whose weights are frozen. This two-step approach has several advantages:\begin{enumerate}
 	\item The original training of M\&Ms-VAE is not perturbed by the critiquing objective.
 	\item $\xi(\cdot)$ is decoupled from the model. It allows more flexibility in its architecture, objective function, and training.
 \end{enumerate}

 We assume that each critique is independent, as in~\cite{luo2020b,wu2019,antognini2020interacting}. At time step $t$, we express the new user preferences~$\btzuct$ with a linear interpolation between the original latent representation $\bzu^0$ and the critique $\bcut$'s representation $\bzuct$. More precisely, we use the gating mechanism of the gated recurrent unit~\cite{chung2014empirical} (we omit the biases to reduce notational clutter):{\small \begin{equation}
 \begin{split}
 	\btzuct = \xi(\bzu^0,\bzuct) = h_2
 \end{split}
\quad\quad\quad\quad\quad\quad
  \begin{split}
    h_2 &= (1-u_1) \cdot n_1 + u_1 \cdot h_1\\
    n_1 &= \textrm{tanh}(W_{in} \bzuct + W_{hn}(r_1 \odot h_1))\\
    u_1 &= \sigma(W_{iu} \bzuct + W_{hz} h_1)\\
    r_1 &= \sigma(W_{ir} \bzuct + W_{hr} h_1)
  \end{split}
\quad\quad\quad\quad
  \begin{split}
    h_1 &= (1-u_0) \cdot n_0 + u_0 \cdot h_0\\
    n_0 &= \textrm{tanh}(W_{in} \bzu^0 + W_{hn}(r_0 \odot h_0))\\
    u_0 &= \sigma(W_{iu} \bzu^0 + W_{hz} h_0)\\
    r_0 &= \sigma(W_{ir} \bzu^0 + W_{hr} h_0)
  \end{split},
\end{equation}} where $W_{ir},W_{iu},W_{in},W_{hr},W_{hz},W_{hn}$, and the bias vectors are the model parameters.

\subsubsection{Training}
Thanks to our assumption and the generalization ability of M\&Ms-VAE, we can learn the weights of the blending module $\xi(\cdot)$ by creating a synthetic dataset based only on the validation set (see Algorithm~\ref{alg_dataset}).~For each user and observed interaction, we randomly sample a keyphrase $\bc$ that is inconsistent with the target item. Then, we calculate the item sets $I^{+\bc}$ that contain the critique and, symmetrically, the item sets $I^{-\bc}$ for those that do not~contain~it.\footnote{In early experiments, we generalized $\xi(\bzu^0,\bzuct)$ to $\xi(\bzu^0,\bzu^1,\dots,\bz_{u}^{t})$ and updated Algorithm~\ref{alg_dataset} accordingly. However, our synthetic dataset cannot cover such a space due to the exponential number of combinations; session-based recommenders require millions of real sessions as training data~\cite{hidasi2015sessionbased,hidasi2018sessionbased}.}

Our final objective is to re-rank items based on the user preferences and the provided critique $\bc$. Recall that M\&Ms-VAE's weights are frozen. Let $\bhru^0$ be the user $u$'s initial predictions $\bhru$ and $\bhru^1$ those inferred from $\btzuct$ after the critique. We express this overall ranking-based objective via two differentiable max-margin objective functions:
\begin{equation}
	\label{eq_ssc}
	\mathcal{L}(\bhRzero,\bhRone,u,\bc,I^{+\bc},I^{-\bc}) = \sum_{i^+ \in I^{+\bc}} \biggl[ \max \Big\{0, h - (\bhr_{u,i^+}^{0} - \bhr_{u,i^+}^{1})\Big\} \biggr]
	 + \sum_{i^- \in I^{-\bc}} \biggl[ \max \Big\{0, h - (\bhr_{u,i^-}^{1} - \bhr_{u,i^-}^{0})\Big\} \biggr],
\end{equation} where $h$ is the margin. Intuitively, $\xi(\cdot)$ is encouraged to create a representation $\btzuct$ from which $p_{\Theta_r}(\cdot)$ gives a lower ranking to the items affected by the critique in the next iteration (i.e., $\bhr_{u,i^+}^{1} < \bhr_{u,i^+}^{0}$) and a higher ranking to the unaffected items (i.e., $\bhr_{u,i^-}^{1} > \bhr_{u,i^-}^{0}$).
Finally, Equation~\ref{eq_ssc} is efficiently parallelizable on both CPUs and GPUs. 

\section{Experiments}

In this section, we proceed to evaluate the proposed M\&Ms-VAE model in order to answer the following questions:
\begin{itemize}
	\item \textbf{RQ 1}: How does M\&Ms-VAE perform in terms of recommendation and explanation performance?
	\item \textbf{RQ 2}: Can M\&Ms-VAE with the self-supervised critiquing objective enable multi-step critiquing?
	\item \textbf{RQ 3}: What is our proposed critiquing algorithm's computational time complexity compared to prior~work? 
	\item \textbf{RQ 4}: How does M\&Ms-VAE perform under weak supervision; how coherent is the joint and cross generation?
\end{itemize}

\subsection{Datasets}
\label{sec_datasets}
We evaluate the quantitative performance of M\&Ms-VAE using four real-world, publicly available datasets: BeerAdvocate~\cite{beer}, Amazon CDs\&Vinyl~\cite{amazon1,amazon2}, Yelp~\cite{yelpdataset}, and HotelRec~\cite{antognini-faltings:2020:LREC1}. Each contains more than 100k reviews with~five-star ratings. For the purpose of Top-N recommendation, we binarize the ratings with a threshold $t > 3.5$. Because~people tend to rate beers and restaurants positively, we set the threshold $t > 4$ and $t > 4.5$, respectively. We split each dataset into~60\%/20\%/20\% for the training, validation, and test sets. Table~\ref{stats_datasets} shows the statistics of the~datasets. All contain complete~observations.
The datasets do not contain preselected keyphrases. Hence, we extract the keyphrases for the explanations and critiquing with the frequency-based processing of~\cite{wu2019,hanze2020}. Some examples are shown in the~appendix. 

\begin{table}[t]
    \centering
   \caption{\label{stats_datasets}Descriptive statistics of the datasets. Coverage shows the ratio of reviews having at least one of the selected keyphrases~(KPs)}
\begin{threeparttable}
\begin{tabular}{@{}l@{}c@{\hspace{2mm}}c@{\hspace{2mm}}c@{\hspace{2mm}}c@{\hspace{2mm}}c@{\hspace{2mm}}c@{\hspace{2mm}}c@{\hspace{3mm}}c@{}c@{}}
\textbf{Dataset} & \textbf{\#Users} & \textbf{\#Items} & \textbf{\#Interactions} & \textbf{Sparsity} & \textbf{\#KPs} & \textbf{KP Coverage} & \textbf{Avg. KPs/Review} & \textbf{AVG. KPs/User}\\
\toprule
Beer & 6,370 & 3,669 & 263,244 & 1.13\% & 75 & 99.27\% & 7.16 & 1,216\\
CDs\&Vinyl & 6,060 & 4,395 & 152,783 & 0.57\% & 40 & 74.59\% & 2.13 & 73\\
Yelp & 9,801 & 4,706 & 140,496 & 0.30\% & 234 & 96.65\% & 7.45 & 300\\
Hotel & 7,044 & 4,874 & 143,612 & 0.42\% & 141 & 99.99\% &17.42 & 419

\end{tabular}
\end{threeparttable}
\end{table}
\subsection{Experimental Settings}
Across experiments, we treat the prior and the likelihood as standard normal and multinomial distributions, respectively. The inference and generative networks consist of a two-layer neural network with a tanh nonlinearity as the activation function between the layers. We normalize the input and use dropout~\cite{srivastava2014dropout}. For learning, we employ the Adam optimizer~\cite{KingmaB14} with AMSGrad~\cite{47409} and a learning rate of $5\cdot10^{-5}$. We anneal linearly the regularization parameter $\beta$ of the Kullback-Leibler terms.
For the baselines, we reused the authors' code and tuning procedure.
We select hyperparameters and architectures for each model by evaluating NDCG on the validation set. We limit the search to a maximum of 100 trials. For critiquing, we tune our blending module on the synthetic dataset with the Falling MAP metric on the validation set, which measures the effect of a critique~\cite{wu2019}. 
For reproducibility purposes, we include additional details and the best hyperparameters in the supplementary~material.

\subsection{RQ 1: How does M\&Ms-VAE perform in terms of recommendation and explanation performance?}
\label{sec_rq1}

\subsubsection{Baselines} \label{sec_rq1_baselines}We compare our proposed M\&Ms-VAE model to the following baseline models. \textbf{POP} returns the most popular items without any kind of personalization. \textbf{AutoRec}~\cite{10.1145/2740908.2742726} is a neural autoencoder-based recommendation system. \textbf{BPR}~\cite{10.5555/1795114.1795167} is a Bayesian personalized ranking model that explicitly optimizes pairwise rankings. \textbf{CDAE}~\cite{10.1145/2835776.2835837} denotes a collaborative denoising autoencoder that is specifically optimized for implicit feedback recommendation tasks. \textbf{NCE-PLRec}~\cite{10.1145/3331184.3331201} represents the linear recommendation projected by noise-contrastive estimation; it augments PLRec with noise-contrasted item embeddings. \textbf{PLRec}~\cite{sedhain2016practical} is the ablation variant of NCE-PLRec without the noise-contrastive estimation. \textbf{PureSVD}~\cite{10.1145/1864708.1864721} denotes a similarity-based recommendation method that constructs a similarity matrix through SVD decomposition of the implicit rating matrix. \textbf{VAE-CF} is the variational autoencoder for collaborative filtering described in Section~\ref{sec_vae}. \textbf{CE-VNCF}~\cite{wu2019} is the extension of the neural collaborative filtering model~\cite{he2017neural} that is augmented with an explanation and a critiquing neural component. Finally, \textbf{CE-VAE}~\cite{luo2020} is a significant improvement over CE-VNCF, and it produces state-of-the-art performance (more details in Section~\ref{sec_cevae}).
For a fair comparison, we encode the user observations in M\&Ms-VAE using solely the inference network $q_{\Phi_r}(\bzu | \bru)$ at test time.

\subsubsection{Top-N Recommendation Performance}
\label{sec_rq1_rec} We report the following five metrics: R-Precision and NDCG; MAP,~Precision, and Recall at different Top-N. The main results are presented in Table~\ref{table_rec_perf}. We make the following key observations. 
\begin{table}[t]
\small
    \centering
\caption{\label{table_rec_perf}Top-N recommendation results of all datasets. \textbf{Bold} and \underline{underline} denote the best and second-best results, respectively. We omit the error bars because the 95\% confidence interval is in 4\textsuperscript{th} digit.}
\begin{threeparttable}
\begin{tabular}{@{}cl@{}
c@{\hspace{1mm}}c@{\hspace{0mm}}c@{\hspace{4mm}}
c@{\hspace{2mm}}c@{\hspace{2mm}}c@{}c@{\hspace{4mm}}
c@{\hspace{2mm}}c@{\hspace{2mm}}c@{}c@{\hspace{4mm}}
c@{\hspace{2mm}}c@{\hspace{2mm}}c@{}}
& & & & & \multicolumn{3}{c}{\textbf{MAP@N}} & & \multicolumn{3}{c}{\textbf{Precision@N}} & & \multicolumn{3}{c}{\textbf{Recall@N}}\\
\cmidrule{6-8}\cmidrule{10-12}\cmidrule{14-16}
& \textbf{Model} & \textbf{R-Precision} & \textbf{NDCG} & & $N=5$ & $N=10$ & $N=20$ & & $N=5$ & $N=10$ & $N=20$ & & $N=5$ & $N=10$ & $N=20$\\
\toprule
\multirow{11.5}{*}{\rotatebox{90}{\textit{Beer}}}
& POP & $0.0307$ & $0.0777$ &  & $0.0388$ & $0.0350$ & $0.0319$ &  & $0.0346$ & $0.0298$ & $0.0279$ &  & $0.0241$ & $0.0408$ & $0.0737$ \\
& AutoRec & $0.0496$ & $0.1140$ &  & $0.0652$ & $0.0591$ & $0.0527$ &  & $0.0574$ & $0.0503$ & $0.0438$ &  & $0.0392$ & $0.0663$ & $0.1129$ \\
& BPR & $0.0520$ & $0.1214$ &  & $0.0646$ & $0.0597$ & $0.0538$ &  & $0.0596$ & $0.0525$ & $0.0449$ &  & $0.0451$ & $0.0744$ & $0.1214$ \\
& CDAE & $0.0414$ & $0.0982$ &  & $0.0504$ & $0.0477$ & $0.0434$ &  & $0.0482$ & $0.0432$ & $0.0368$ &  & $0.0330$ & $0.0576$ & $0.0969$ \\
& NCE-PLRec & $0.0501$ & $0.1151$ &  & $0.0643$ & $0.0594$ & $0.0532$ &  & $0.0589$ & $0.0518$ & $0.0440$ &  & $0.0418$ & $0.0714$ & $0.1177$ \\
& PLRec & $0.0497$ & $0.1113$ &  & $0.0655$ & $0.0599$ & $0.0532$ &  & $0.0590$ & $0.0515$ & $0.0431$ &  & $0.0421$ & $0.0704$ & $0.1127$ \\
& PureSVD & $0.0450$ & $0.1052$ &  & $0.0479$ & $0.0473$ & $0.0446$ &  & $0.0493$ & $0.0455$ & $0.0396$ &  & $0.0391$ & $0.0689$ & $0.1131$ \\
& VAE-CF & $\underline{0.0538}$ & $\underline{0.1275}$ &  & $0.0642$ & $0.0594$ & $0.0536$ &  & $0.0595$ & $0.0525$ & $0.0448$ &  & $\underline{0.0473}$ & $\underline{0.0808}$ & $\underline{0.1327}$ \\
& CE-VAE & $0.0520$ & $0.1215$ &  & $\underline{0.0675}$ & $\underline{0.0618}$ & $\underline{0.0555}$ &  & $\underline{0.0620}$ & $\underline{0.0536}$ & $\underline{0.0461}$ &  & $0.0442$ & $0.0737$ & $0.1255$ \\
& CE-VNCF & $0.0440$ & $0.1099$ &  & $0.0546$ & $0.0512$ & $0.0472$ &  & $0.0504$ & $0.0465$ & $0.0411$ &  & $0.0353$ & $0.0635$ & $0.1116$ \\
& M\&Ms-VAE (Ours) & $\textbf{0.0545}$ & $\textbf{0.1307}$ &  & $\textbf{0.0706}$ & $\textbf{0.0650}$ & $\textbf{0.0580}$ &  & $\textbf{0.0649}$ & $\textbf{0.0563}$ & $\textbf{0.0473}$ &  & $\textbf{0.0492}$ & $\textbf{0.0833}$ & $\textbf{0.1349}$ \\
\bottomrule
\multirow{11.5}{*}{\rotatebox{90}{\textit{CDs\&Vinyl}}}
& POP & $0.0088$ & $0.0265$ &  & $0.0108$ & $0.0102$ & $0.0095$ &  & $0.0098$ & $0.0095$ & $0.0082$ &  & $0.0088$ & $0.0182$ & $0.0327$ \\
& AutoRec & $0.0227$ & $0.0537$ &  & $0.0284$ & $0.0257$ & $0.0220$ &  & $0.0255$ & $0.0213$ & $0.0165$ &  & $0.0254$ & $0.0418$ & $0.0627$ \\
& BPR & $0.0632$ & $0.1516$ &  & $0.0724$ & $0.0639$ & $0.0543$ &  & $0.0640$ & $0.0513$ & $0.0408$ &  & $0.0807$ & $0.1263$ & $0.1939$ \\
& CDAE & $0.0135$ & $0.0365$ &  & $0.0173$ & $0.0158$ & $0.0141$ &  & $0.0152$ & $0.0136$ & $0.0116$ &  & $0.0143$ & $0.0262$ & $0.0451$ \\
& NCE-PLRec & $0.0749$ & $\underline{0.1739}$ &  & $0.0728$ & $0.0678$ & $0.0586$ &  & $0.0698$ & $0.0584$ & $0.0441$ &  & $\textbf{0.1010}$ & $\textbf{0.1608}$ & $\underline{0.2308}$ \\
& PLRec & $\underline{0.0760}$ & $0.1626$ &  & $\textbf{0.0889}$ & $\underline{0.0777}$ & $\underline{0.0642}$ &  & $\underline{0.0773}$ & $\underline{0.0608}$ & $\underline{0.0444}$ &  & $0.0960$ & $0.1461$ & $0.2025$ \\
& PureSVD & $0.0652$ & $0.1551$ &  & $0.0570$ & $0.0565$ & $0.0509$ &  & $0.0612$ & $0.0527$ & $0.0405$ &  & $0.0914$ & $0.1486$ & $0.2149$ \\
& VAE-CF & $0.0638$ & $0.1699$ &  & $0.0540$ & $0.0554$ & $0.0517$ &  & $0.0600$ & $0.0540$ & $0.0440$ &  & $0.0949$ & $\underline{0.1593}$ & $\textbf{0.2381}$ \\
& CE-VAE & $0.0708$ & $0.1532$ &  & $0.0816$ & $0.0711$ & $0.0588$ &  & $0.0715$ & $0.0555$ & $0.0411$ &  & $0.0903$ & $0.1357$ & $0.1937$ \\
& CE-VNCF & $0.0654$ & $0.1524$ &  & $0.0746$ & $0.0662$ & $0.0560$ &  & $0.0663$ & $0.0534$ & $0.0411$ &  & $0.0829$ & $0.1299$ & $0.1931$ \\
& M\&Ms-VAE (Ours) & $\textbf{0.0801}$ & $\textbf{0.1765}$ &  & $\underline{0.0885}$ & $\textbf{0.0784}$ & $\textbf{0.0660}$ &  & $\textbf{0.0779}$ & $\textbf{0.0628}$ & $\textbf{0.0482}$ &  & $\underline{0.0983}$ & $0.1529$ & $0.2263$ \\
\bottomrule
\multirow{11.5}{*}{\rotatebox{90}{\textit{Yelp}}}
& POP & $0.0026$ & $0.0129$ &  & $0.0024$ & $0.0026$ & $0.0026$ &  & $0.0028$ & $0.0028$ & $0.0025$ &  & $0.0042$ & $0.0087$ & $0.0151$ \\
& AutoRec & $0.0034$ & $0.0133$ &  & $0.0032$ & $0.0030$ & $0.0028$ &  & $0.0027$ & $0.0027$ & $0.0025$ &  & $0.0038$ & $0.0081$ & $0.0153$ \\
& BPR & $0.0160$ & $0.0609$ &  & $0.0168$ & $0.0156$ & $0.0143$ &  & $0.0156$ & $0.0140$ & $0.0122$ &  & $0.0236$ & $0.0435$ & $0.0748$ \\
& CDAE & $0.0028$ & $0.0135$ &  & $0.0027$ & $0.0028$ & $0.0027$ &  & $0.0030$ & $0.0027$ & $0.0026$ &  & $0.0044$ & $0.0084$ & $0.0161$ \\
& NCE-PLRec & $0.0197$ & $0.0739$ &  & $0.0220$ & $0.0200$ & $0.0177$ &  & $0.0198$ & $0.0169$ & $0.0143$ &  & $0.0300$ & $0.0505$ & $0.0864$ \\
& PLRec & $0.0191$ & $0.0703$ &  & $0.0207$ & $0.0189$ & $0.0171$ &  & $0.0185$ & $0.0166$ & $0.0143$ &  & $0.0291$ & $0.0513$ & $0.0866$ \\
& PureSVD & $\underline{0.0253}$ & $\underline{0.0825}$ &  & $\underline{0.0279}$ & $\underline{0.0249}$ & $\underline{0.0217}$ &  & $\underline{0.0240}$ & $\underline{0.0206}$ & $\underline{0.0173}$ &  & $\underline{0.0357}$ & $\underline{0.0597}$ & $\underline{0.1008}$ \\
& VAE-CF & $0.0214$ & $0.0801$ &  & $0.0232$ & $0.0216$ & $0.0195$ &  & $0.0214$ & $0.0192$ & $0.0163$ &  & $0.0319$ & $0.0589$ & $0.0995$ \\
& CE-VAE & $0.0136$ & $0.0533$ &  & $0.0143$ & $0.0132$ & $0.0121$ &  & $0.0125$ & $0.0119$ & $0.0104$ &  & $0.0197$ & $0.0367$ & $0.0636$ \\
& CE-VNCF & $0.0166$ & $0.0693$ &  & $0.0175$ & $0.0167$ & $0.0157$ &  & $0.0165$ & $0.0154$ & $0.0144$ &  & $0.0251$ & $0.0467$ & $0.0889$ \\
& M\&Ms-VAE (Ours) & $\textbf{0.0264}$ & $\textbf{0.0909}$ &  & $\textbf{0.0284}$ & $\textbf{0.0261}$ & $\textbf{0.0231}$ &  & $\textbf{0.0260}$ & $\textbf{0.0223}$ & $\textbf{0.0188}$ &  & $\textbf{0.0395}$ & $\textbf{0.0682}$ & $\textbf{0.1154}$ \\
\bottomrule
\multirow{11.5}{*}{\rotatebox{90}{\textit{Hotel}}}
& POP & $0.0047$ & $0.0188$ &  & $0.0049$ & $0.0047$ & $0.0042$ &  & $0.0047$ & $0.0043$ & $0.0036$ &  & $0.0054$ & $0.0098$ & $0.0167$ \\
& AutoRec & $0.0051$ & $0.0193$ &  & $0.0053$ & $0.0050$ & $0.0044$ &  & $0.0052$ & $0.0042$ & $0.0037$ &  & $0.0061$ & $0.0097$ & $0.0169$ \\
& BPR & $0.0181$ & $0.0623$ &  & $0.0198$ & $0.0185$ & $0.0169$ &  & $0.0183$ & $0.0168$ & $0.0146$ &  & $0.0219$ & $0.0409$ & $0.0713$ \\
& CDAE & $0.0050$ & $0.0190$ &  & $0.0054$ & $0.0049$ & $0.0044$ &  & $0.0050$ & $0.0043$ & $0.0037$ &  & $0.0057$ & $0.0098$ & $0.0172$ \\
& NCE-PLRec & $0.0229$ & $0.0684$ &  & $0.0244$ & $0.0226$ & $0.0200$ &  & $0.0228$ & $0.0195$ & $0.0160$ &  & $0.0283$ & $0.0484$ & $0.0785$ \\
& PLRec & $0.0242$ & $0.0664$ &  & $0.0265$ & $0.0234$ & $0.0201$ &  & $0.0228$ & $0.0190$ & $0.0155$ &  & $0.0284$ & $0.0466$ & $0.0758$ \\
& PureSVD & $0.0179$ & $0.0541$ &  & $0.0193$ & $0.0173$ & $0.0152$ &  & $0.0169$ & $0.0145$ & $0.0121$ &  & $0.0206$ & $0.0357$ & $0.0594$ \\
& VAE-CF & $\underline{0.0243}$ & $\underline{0.0755}$ &  & $\underline{0.0267}$ & $\underline{0.0241}$ & $\underline{0.0213}$ &  & $\underline{0.0238}$ & $\underline{0.0206}$ & $\underline{0.0171}$ &  & $\underline{0.0295}$ & $\underline{0.0511}$ & $\underline{0.0848}$ \\
& CE-VAE & $0.0147$ & $0.0538$ &  & $0.0151$ & $0.0146$ & $0.0136$ &  & $0.0148$ & $0.0137$ & $0.0122$ &  & $0.0184$ & $0.0334$ & $0.0595$ \\
& CE-VNCF & $0.0165$ & $0.0575$ &  & $0.0180$ & $0.0166$ & $0.0152$ &  & $0.0159$ & $0.0149$ & $0.0129$ &  & $0.0200$ & $0.0370$ & $0.0635$ \\
& M\&Ms-VAE (Ours) & $\textbf{0.0272}$ & $\textbf{0.0804}$ &  & $\textbf{0.0290}$ & $\textbf{0.0265}$ & $\textbf{0.0235}$ &  & $\textbf{0.0260}$ & $\textbf{0.0227}$ & $\textbf{0.0189}$ &  & $\textbf{0.0314}$ & $\textbf{0.0555}$ & $\textbf{0.0928}$
\end{tabular}
\end{threeparttable}
\end{table}
Overall, our proposed M\&Ms-VAE model shows the best recommendation performance for all metrics on three datasets and nearly all metrics on the CDs\&Vinyl dataset.
Compared to the original VAE recommender~(VAE-CF), M\&Ms-VAE achieves an improvement of 13\% on average. We conjecture that the additional loss terms (i.e.,~$ELBO(\bru, \bku)$ and $ELBO(\bku)$) help to generate better user representations by leveraging both the user preferences and keyphrase usage with the mixture of experts.

M\&Ms-VAE also significantly outperforms CE-VAE on the Yelp and Hotel datasets (by a factor of 1.9 and 1.7, respectively) and achieves an average improvement of 9\% on the Beer and CDs\&Vinyl datasets. We remark the same trend with CE-VNCF. These results emphasize the noise introduced in CE-VAE and CE-VNCF during training when learning the mapping between the keyphrases and the latent space. This is even more pronounced with a large number of keyphrases (i.e., over 100). In contrast, M\&Ms-VAE is more robust thanks to our factorization and training strategy.

Interestingly, PureSVD exhibits the second-best performance on the CDs\&Vinyl and Yelp datasets. This shows that classic algorithms often remain competitive with state-of-the-art VAE-based recommender systems.
\begin{table}[!t]
\small
    \centering
\caption{\label{table_exp_perf}Top-K keyphrase explanation quality of all datasets. \textbf{Bold} and \underline{underline} denote the best and second-best results, respectively. We omit the error bars because the 95\% confidence interval is in 4\textsuperscript{th} digit.}
\begin{threeparttable}
\begin{tabular}{@{}c@{\hspace{2mm}}l
@{\hspace{1mm}}c@{\hspace{2mm}}c@{\hspace{2mm}}c@{}c@{\hspace{3.5mm}}
c@{\hspace{2mm}}c@{\hspace{2mm}}c@{}c@{\hspace{3.5mm}}
c@{\hspace{2mm}}c@{\hspace{2mm}}c@{}c@{\hspace{3.5mm}}
c@{\hspace{2mm}}c@{\hspace{2mm}}c@{}}
& & \multicolumn{3}{c}{\textbf{NDCG@K}} & & \multicolumn{3}{c}{\textbf{MAP@K}} & & \multicolumn{3}{c}{\textbf{Precision@K}} & & \multicolumn{3}{c}{\textbf{Recall@K}}\\
\cmidrule{3-5}\cmidrule{7-9}\cmidrule{11-13}\cmidrule{15-17}
& \textbf{Model} & $K=5$ & $K=10$ & $K=20$ & & $K=5$ & $K=10$ & $K=20$ & & $K=5$ & $K=10$ & $K=20$ & & $K=5$ & $K=10$ & $K=20$\\
\toprule
\multirow{5.5}{*}{\rotatebox{90}{\textit{Beer}}}
& UserPop & $0.0550$ & $0.0852$ & $0.1484$ &  & $0.0812$ & $0.0795$ & $0.0824$ &  & $0.0782$ & $0.0799$ & $0.0933$ &  & $0.0446$ & $0.0913$ & $0.2186$ \\
& ItemPop & $0.0511$ & $0.0807$ & $0.1428$ &  & $0.0767$ & $0.0749$ & $0.0777$ &  & $0.0697$ & $0.0740$ & $0.0895$ &  & $0.0402$ & $0.0856$ & $0.2107$ \\
& CE-VAE & $\underline{0.2803}$ & $\underline{0.4104}$ & $\underline{0.5916}$ &  & $\underline{0.9418}$ & $\underline{0.9168}$ & $\underline{0.8820}$ &  & $\underline{0.9186}$ & $\underline{0.8760}$ & $\underline{0.8186}$ &  & $0.1448$ & $0.2683$ & $0.4829$ \\
& CE-VNCF & $0.2390$ & $0.3221$ & $0.4080$ &  & $0.3414$ & $0.3117$ & $0.2690$ &  & $0.3145$ & $0.2633$ & $0.2026$ &  & $\textbf{0.1966}$ & $\textbf{0.3263}$ & $\textbf{0.4962}$ \\
& M\&Ms-VAE (Ours) & $\textbf{0.2817}$ & $\textbf{0.4147}$ & $\textbf{0.5960}$ &  & $\textbf{0.9463}$ & $\textbf{0.9237}$ & $\textbf{0.8885}$ &  & $\textbf{0.9243}$ & $\textbf{0.8861}$ & $\textbf{0.8256}$ &  & $0.1457$ & $0.2722$ & $\underline{0.4869}$ \\
\bottomrule
\multirow{5.5}{*}{\rotatebox{90}{\textit{CDs\&Vinyl}}}
& UserPop & $0.1028$ & $0.1285$ & $0.1869$ &  & $0.0910$ & $0.0807$ & $0.0700$ &  & $0.0860$ & $0.0636$ & $0.0595$ &  & $0.1157$ & $0.1704$ & $0.3392$ \\
& ItemPop & $0.1109$ & $0.1357$ & $0.1935$ &  & $0.0928$ & $0.0833$ & $0.0716$ &  & $0.0929$ & $0.0657$ & $0.0606$ &  & $0.1288$ & $0.1825$ & $0.3499$ \\
& CE-VAE & $\underline{0.5243}$ & $\underline{0.6468}$ & $\underline{0.7454}$ &  & $\underline{0.6441}$ & $\underline{0.5609}$ & $\underline{0.4575}$ &  & $\underline{0.5564}$ & $\underline{0.4324}$ & $\underline{0.3040}$ &  & $0.4795$ & $\underline{0.6808}$ & $\underline{0.8757}$ \\
& CE-VNCF & $0.4590$ & $0.5338$ & $0.5860$ &  & $0.3657$ & $0.2981$ & $0.2258$ &  & $0.2893$ & $0.2010$ & $0.1260$ &  & $\textbf{0.5081}$ & $0.6698$ & $0.8127$ \\
& M\&Ms-VAE (Ours) & $\textbf{0.5447}$ & $\textbf{0.6659}$ & $\textbf{0.7628}$ &  & $\textbf{0.6648}$ & $\textbf{0.5779}$ & $\textbf{0.4700}$ &  & $\textbf{0.5777}$ & $\textbf{0.4441}$ & $\textbf{0.3091}$ &  & $\underline{0.5015}$ & $\textbf{0.6996}$ & $\textbf{0.8894}$ \\
\bottomrule
\multirow{5.5}{*}{\rotatebox{90}{\textit{Yelp}}}
& UserPop & $0.0007$ & $0.0009$ & $0.0066$ &  & $0.0009$ & $0.0010$ & $0.0016$ &  & $0.0013$ & $0.0009$ & $0.0061$ &  & $0.0007$ & $0.0011$ & $0.0129$ \\
& ItemPop & $0.0008$ & $0.0011$ & $0.0073$ &  & $0.0009$ & $0.0011$ & $0.0018$ &  & $0.0015$ & $0.0011$ & $0.0065$ &  & $0.0009$ & $0.0013$ & $0.0149$ \\
& CE-VAE & $\underline{0.1935}$ & $\underline{0.2763}$ & $\underline{0.3803}$ &  & $\underline{0.6653}$ & $\underline{0.6356}$ & $\underline{0.5916}$ &  & $\underline{0.6363}$ & $\underline{0.5876}$ & $\underline{0.5181}$ &  & $\underline{0.1017}$ & $\underline{0.1819}$ & $\underline{0.3080}$ \\
& CE-VNCF & $0.0883$ & $0.1164$ & $0.1505$ &  & $0.1195$ & $0.1052$ & $0.0901$ &  & $0.1023$ & $0.0848$ & $0.0690$ &  & $0.0779$ & $0.1270$ & $0.2027$ \\
& M\&Ms-VAE (Ours) & $\textbf{0.1949}$ & $\textbf{0.2787}$ & $\textbf{0.3837}$ &  & $\textbf{0.6738}$ & $\textbf{0.6428}$ & $\textbf{0.5976}$ &  & $\textbf{0.6434}$ & $\textbf{0.5935}$ & $\textbf{0.5229}$ &  & $\textbf{0.1019}$ & $\textbf{0.1834}$ & $\textbf{0.3107}$ \\
\bottomrule
\multirow{5.5}{*}{\rotatebox{90}{\textit{Hotel}}}
& UserPop & $0.0436$ & $0.0681$ & $0.1059$ &  & $0.1091$ & $0.1059$ & $0.1054$ &  & $0.1018$ & $0.1036$ & $0.1050$ &  & $0.0265$ & $0.0557$ & $0.1120$ \\
& ItemPop & $0.0483$ & $0.0756$ & $0.1152$ &  & $0.1159$ & $0.1137$ & $0.1124$ &  & $0.1108$ & $0.1131$ & $0.1111$ &  & $0.0303$ & $0.0633$ & $0.1237$ \\
& CE-VAE & $\underline{0.2425}$ & $\underline{0.3521}$ & $\underline{0.4984}$ &  & $\underline{0.9389}$ & $\underline{0.9113}$ & $\underline{0.8638}$ &  & $\underline{0.9209}$ & $\underline{0.8629}$ & $\underline{0.7831}$ &  & $0.1153$ & $\underline{0.2105}$ & $\underline{0.3704}$ \\
& CE-VNCF & $0.1873$ & $0.2527$ & $0.3280$ &  & $0.3975$ & $0.3671$ & $0.3230$ &  & $0.3732$ & $0.3154$ & $0.2558$ &  & $\textbf{0.1252}$ & $0.2060$ & $0.3230$ \\
& M\&Ms-VAE (Ours) & $\textbf{0.2500}$ & $\textbf{0.3595}$ & $\textbf{0.5054}$ &  & $\textbf{0.9752}$ & $\textbf{0.9393}$ & $\textbf{0.8829}$ &  & $\textbf{0.9498}$ & $\textbf{0.8776}$ & $\textbf{0.7895}$ &  & $\underline{0.1182}$ & $\textbf{0.2131}$ & $\textbf{0.3726}$
\end{tabular}
\end{threeparttable}
\end{table}

\subsubsection{Top-K Explanation Performance}
\label{sec_rq1_exp} We also compare M\&Ms-VAE with the user and item-popularity baselines~\cite{wu2019} that predict the explanation through counting and ranking the frequency of keyphrases for the users (symmetrically, the items) in the training set. Among the recommender baselines, only CE-VAE and CE-VNCF produce an explanation alongside the recommendation. We report the following metrics: NDCG, MAP, Precision, and Recall at different~Top-K. 

Table~\ref{table_exp_perf} contains the main results. Both popularity baselines clearly underperform, showing that the task is not trivial (see Table~\ref{stats_datasets} for the number of keyphrases per dataset). Remarkably, the proposed M\&Ms-VAE model significantly outperforms the CE-VNCF baseline by a factor of 2.5 on average and by approximately by 3.5 on MAP and Precision. 
Moreover, M\&Ms-VAE retrieves 89\% percent of relevant keyphrases within the Top-20 explanations on the CDs\&Vinyl~dataset.

We observe that CE-VAE performs similarly to M\&Ms-VAE but still slightly underperforms by approximately 2\% on average. Finally, we note that CE-VNCF achieves the best results in terms of Recall for the Beer dataset and Recall@5 for the CDs\&Vinyl and Hotel datasets. Nevertheless, as seen in the recommendation performance, CE-VNCF significantly underperforms, highlighting the trade-off between recommendation and explanation.

\subsection{RQ 2: Can M\&Ms-VAE with the self-supervised critiquing objective enable multi-step critiquing?}\label{sec_rq2}

\subsubsection{Baselines}
\textbf{UAC}~\cite{luo2020b} denotes uniform average critiquing, in which the user embedding and all critique embeddings are averaged uniformly. \textbf{BAC}~\cite{luo2020b} is balanced average critiquing. It first averages the critique embeddings and then averages them again with the initial user embedding. \textbf{CE-VAE} and \textbf{CE-VNCF} were introduced in Section~\ref{sec_rq1_baselines}. During training, both learn an inverse feedback loop between a critique and the latent space. At inference, they average the original user embedding with the critique embedding. \textbf{LLC-Score}~\cite{luo2020b} and \textbf{LLC-Rank}~\cite{hanze2020} first extend the PLRec recommender system~\cite{sedhain2016practical} to co-embed keyphrases into the user embedding space with a linear regression. Afterwards, the models apply a weighted average between the initial user embedding and each critique embedding;~the weights are optimized in a linear programming formulation; LLC-Score uses a max-margin scoring-based objective and LLC-Rank a ranking-based objective. To limit computational complexity, the authors limit the number~of constraints to the top-100 rated items.
For a fair comparison, we also consider in M\&Ms-VAE the top-100 rated items~meeting the criteria for $I^{+\bc}$ and $I^{-\bc}$ for each critique $\bc$, although the computational time remains identical using the full sets.

\subsubsection{User Simulation} Similarly to prior work~\cite{luo2020b,hanze2020}, we conduct a user simulation to asses each model's performance in a multi-step conversational recommendation scenario. Concretely, the simulation considers all users and follows Algorithm~\ref{alg_dataset} with the following differences: \begin{enumerate*}
 \item we track the conversational interaction session of simulated users by selecting all target items from their \textbf{test} set,
 \item the maximum allowed critiquing iterations is set to 10, and, 
 \item the conversation stops if the target item appears within the top-N recommendations on that iteration.
\end{enumerate*}

As in~\cite{hanze2020}, we simulate a variety of user-critiquing styles. For each, the candidate keyphrases to critique are inconsistent (i.e., disjoint) with the target item's known keyphrase list. We experiment with the following three~methodologies: \begin{enumerate}
	\item \textbf{Random}: we assume the user randomly chooses a keyphrase to critique.
	\item \textbf{Pop}: we assume the user selects a keyphrase to critique according to the general keyphrase popularity. 
	\item \textbf{Diff}: we assume the user critiques a keyphrase that deviates the most from the known target item description. To do so, we compare the top recommended items' keyphrase frequency to the target item's keyphrases and select the keyphrase with the largest frequency differential.
\end{enumerate}

\subsubsection{Multi-Step Critiquing Performance} Following~\cite{luo2020b,hanze2020}, we asses the models over all users and all items on~the test set using two metrics: the average success rate and session length. The former is the percentage of target items that successfully reach a rank within the Top-N, and the latter is the average length of these sessions (with a limit of 10 iterations). 
In our~experiment framework, for each user and target item, we sample alongside 299 unseen items. 

\begin{figure*}[!t]
\centering
\includegraphics[width=\textwidth,height=6in]{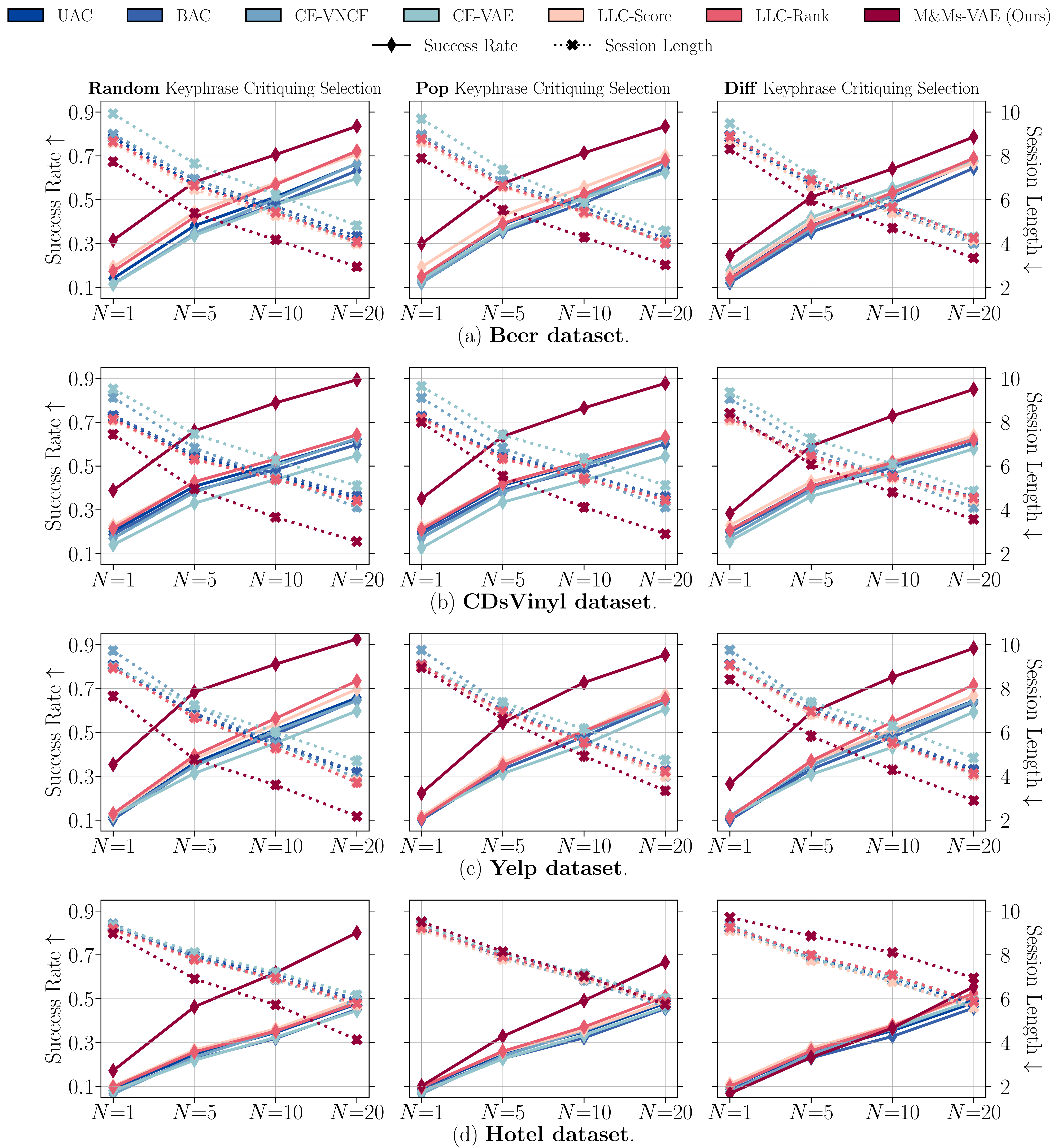}
\caption{\label{fig_multi_crit}Multi-step critiquing performance after 10 turns of conversation. For each dataset and keyphrase critiquing selection method, we report the average success rate (left y-axis) and session length (right y-axis) at different Top-N with 95\% confidence interval.}
\end{figure*}

The results for each dataset and each keyphrase critiquing selection method are depicted in~Figure~\ref{fig_multi_crit}.
Overall, all models' performance is generally better on the Beer and CDs\&Vinyl datasets due to the higher density in terms of~the number of interactions.
Generally, all models tend to find the target item within the Top-20 in more than half the time and under six turns. This highlights that in practice, users are likely to find the desired item in a limited amount of~time.

\begin{figure*}[!t]
\centering
\begin{subfigure}[t]{0.5\textwidth}
    \centering
    \includegraphics[width=0.925\textwidth,height=2.0396in]{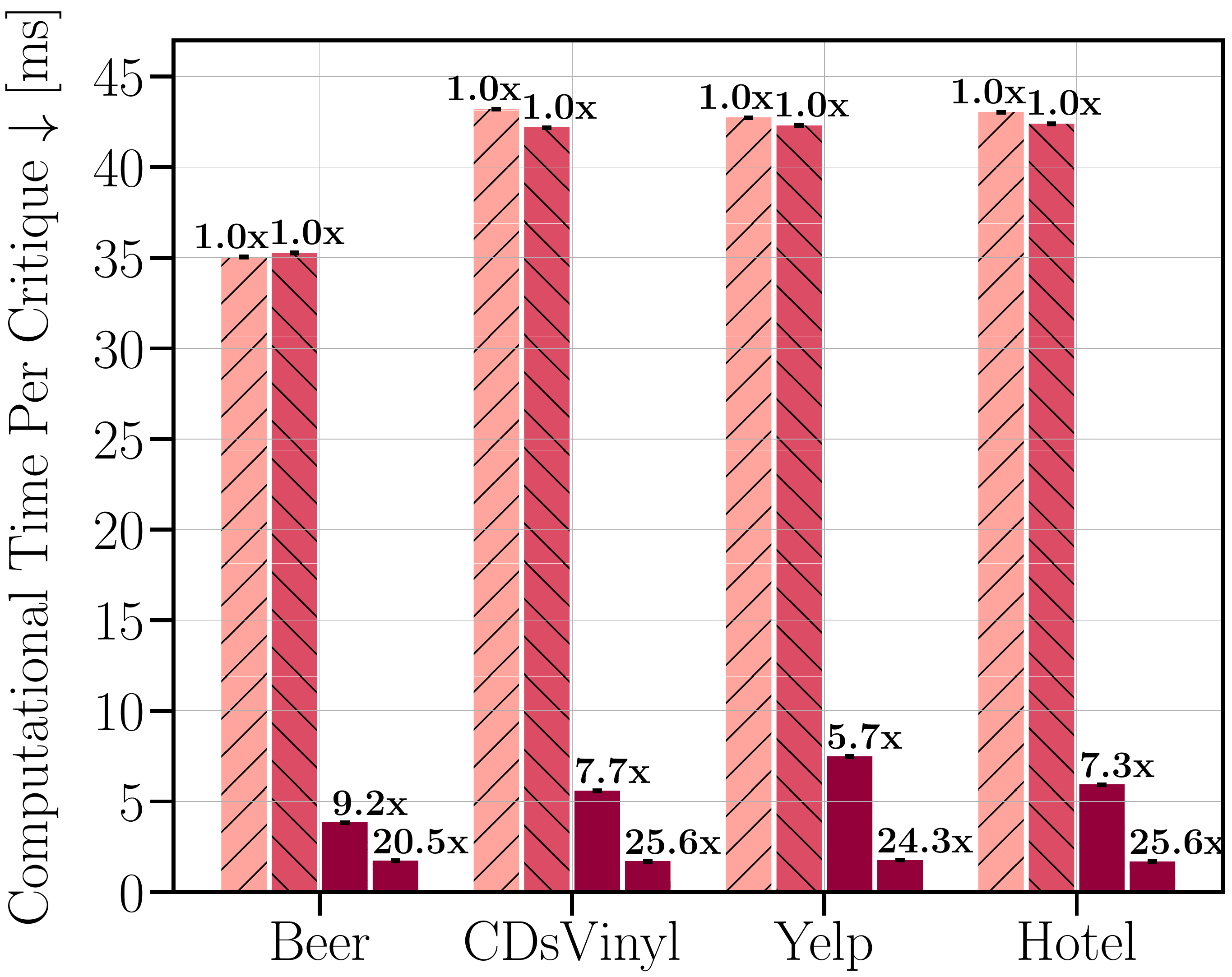}
    \caption{\label{fig_time_analysis_crit}Average computational time of a single critique.}
\end{subfigure}
~
\begin{subfigure}[t]{0.5\textwidth}
    \centering
    \includegraphics[width=0.95\textwidth,height=1.985in]{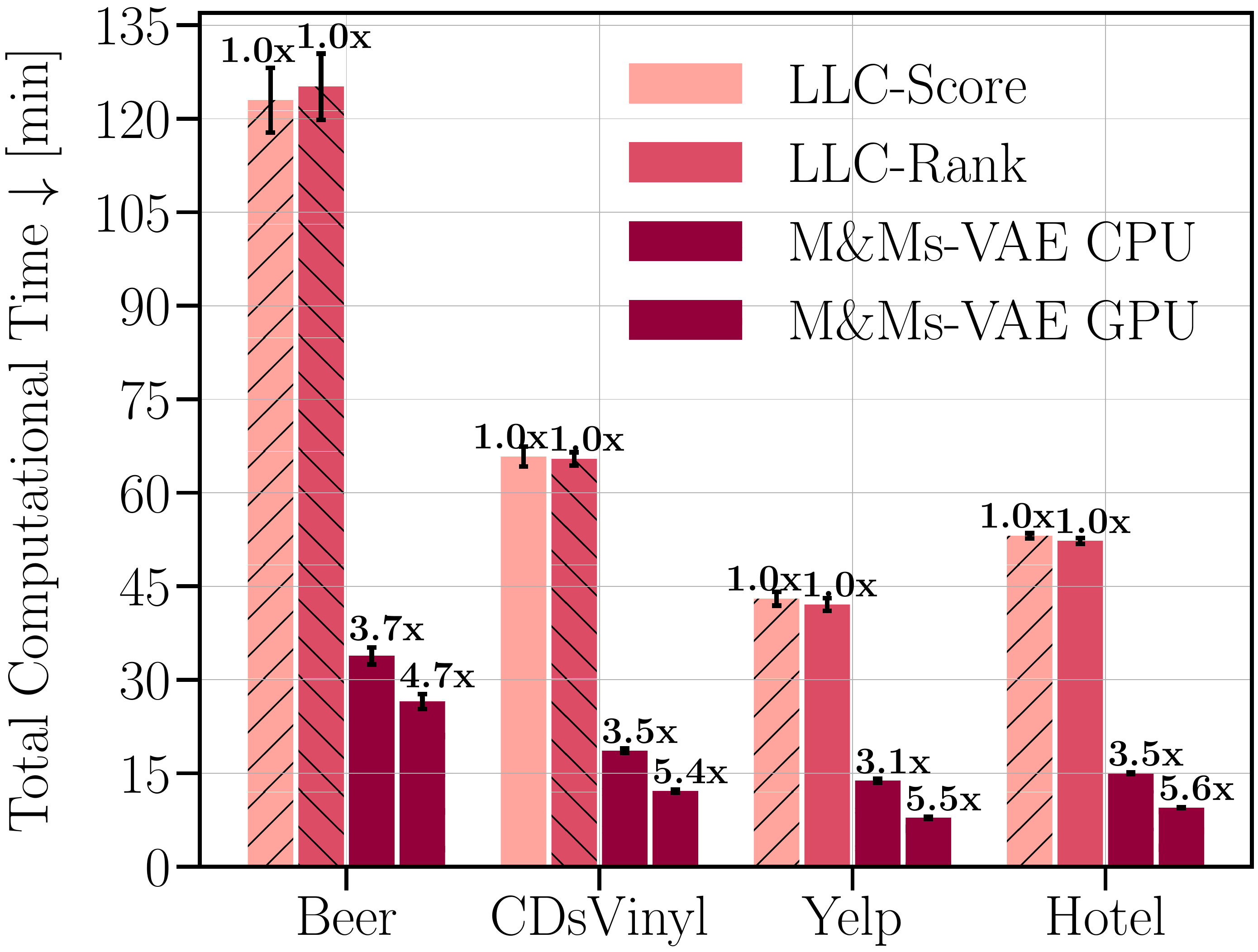}
    \caption{\label{fig_time_analysis_total}Average total computational time of the simulation.}
\end{subfigure}
\caption{\label{fig_time_analysis}Average time consumed for completing 10 runs for 1,000-user simulation after ten turns of conversation with 95\% confidence intervals. LLC-Score and LLC-Rank cannot leverage GPUs; we thus report the performance of M\&Ms-VAE on CPU and GPU. Additionally, we report the inference speedup compared to the slowest model.}
\end{figure*}

Impressively, M\&Ms-VAE clearly outperforms all the baselines on both metrics on the Beer, CDs\&Vinyl, and Yelp datasets. On the Hotel dataset, the success rate is significantly higher for the Random and Pop cases and similar for the Diff case, whereas the session length is higher for the Pop and Diff selection methods. This is unsurprising, because the blending module is trained only once on the random keyphrase critiquing~selection (i.e., no assumption on the user's behavior).\\
\indent Although the simple self-supervision objective in M\&Ms-VAE mimics only one-step random critiquing, we remark that the training strategy generalizes for multi-step critiquing and other keyphrase critiquing selection as well. This shows that M\&Ms-VAE efficiently embeds the critique, thanks to the multimodal modeling.

We observe that the simple UAC and BAC methods perform similarly or better than CE-VNCF and CE-VAE. However, they are outperformed by LLC-Score, LLC-Rank, and M\&Ms-VAE. These results confirm our observation~in Section~\ref{sec_cevae} that the critiquing objective introduces noise during training and does not accurately reflect the critiquing~mechanism.\\
\indent Finally, we note that LLC-Score performs similarly to LLC-Rank in most cases.  When compared to M\&Ms-VAE, both significantly underperform on both metrics in 10 out of 12 cases. This highlights the effectiveness of our proposed critiquing algorithm compared to linear aggregation methods.

\subsection{RQ 3: What is the critiquing computational time complexity for M\&Ms-VAE compared to prior~work?}\label{sec_rq3}
Now, we aim to empirically determine how the critiquing in M\&Ms-VAE compares to the best baselines in Section~\ref{sec_rq2}, LLC-Score and LLC-Rank, in terms of computational time. Because the baselines can not leverage the GPU due to their optimization framework, we also run M\&Ms-VAE on the CPU. We follow the same experiment settings as in Section~\ref{sec_rq2} and limit ourselves to 1,000 users and the Random keyphrase critiquing selection method. All models process exactly 10 critiques for each user-item pair. We employ a machine with a 2.5GHz 24-core CPU, Titan X GPU, and 256GB~memory.

Figure~\ref{fig_time_analysis} shows the average computational time over 10 runs. Particularly, the Figure~\ref{fig_time_analysis_crit} denotes the critiquing computational time in milliseconds, and Figure~\ref{fig_time_analysis_total} the overall simulation time in minutes. Impressively, we observe that the critiquing in M\&Ms-VAE's  is approximately 7.5x faster on CPU and up to 25.6x on the GPU than LLC-Score and LLC-Rank. This shows that once the critiquing module of M\&Ms-VAE is trained, which takes less than five minutes on the machine, the model achieves a lower latency (batch size of one). In Figure~\ref{fig_time_analysis_total}, the overall simulation in M\&Ms-VAE is at least 3.1x faster on CPU and approximately 5.3x faster on GPU. Finally, in real-life applications, we could leverage multiple users' critiques simultaneously and increase the throughput by considering a larger batch size.

\subsection{RQ4: How does M\&Ms-VAE perform under weak supervision; is the joint \& cross inference~coherent?}
\label{seq_rq4}
We first quantify the coherence of joint and cross generations of our M\&Ms-VAE model. We denote~three cases at test time: \begin{enumerate*}
 \item only the user's interactions are used, and the encoder is $q_{\Phi_r}(\bzu | \bru)$;
 \item only the user's keyphrase~preference is used, and the encoder is $q_{\Phi_k}(\bzu | \bku)$; and 
 \item both used with the encoder $q_\Phi(\bzu | \bru, \bku)$.
 \end{enumerate*} Second, we simulate~incomplete supervision by randomly selecting a subset of the training with fully observed samples. The other one~is split into two even parts: the first includes only observed $\bru$ and the second $\bku$. We retain the models and settings of~Section~\ref{sec_rq1}.\\
\begin{figure*}[!t]
\centering
\includegraphics[width=1.0\textwidth]{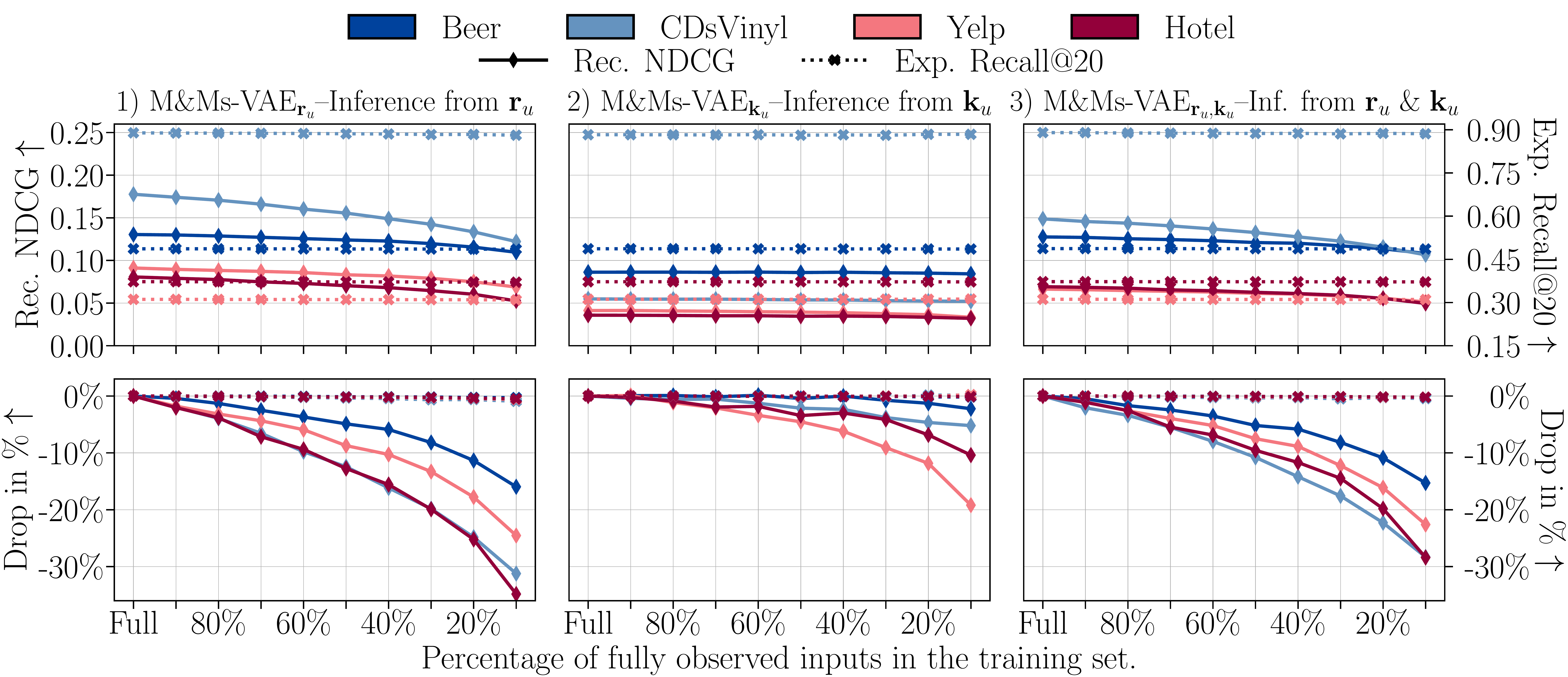}
\caption{\label{fig_rec_exp_mm_perf}Recommendation (left y-axis) and explanation (right y-axis) results averaged on 5 runs with different combinations of~mod-alities observed at inference. The x-axis denotes the percentage of fully observed inputs during training; the rest is partially observed.}
\end{figure*}
\indent Figure~\ref{fig_rec_exp_mm_perf} shows the results averaged on five runs for the three cases at different levels of supervision. The top row presents the explanation and recommendation performance in terms of NDCG and Recall@20 (the model behavior is consistent across the other metrics). On the full datasets, we note that the explanation performance is comparable across the three variants and higher than those of the baselines in Table~\ref{table_exp_perf}. 
Regarding the recommendation performance, M\&Ms-VAE\textsubscript{$\mathbf{r}_u$} obtains the best results, followed closely by M\&Ms-VAE\textsubscript{$\mathbf{r}_u,\mathbf{k}_u$}, and M\&Ms-VAE\textsubscript{$\mathbf{k}_u$} clearly underperforms. This aligns with the observations in~\cite{10.1145/3397271.3401281,10.1145/3178876.3186070}: recommender systems are limited if they use only review~text as input, and not all reviews can be useful. Nevertheless, compared to  Table~\ref{table_rec_perf}, M\&Ms-VAE\textsubscript{$\mathbf{k}_u$} always achieves better recommendation performance than the popularity baseline, and it performs better than AutoRec and CDAE on~two~datasets.

The bottom row of Figure~\ref{fig_rec_exp_mm_perf} show the relative drop in performance on both metrics. The explanation performance seems unaffected by the sparsity, showing that the explanation task remains simple in comparison with the recommendation task. Remarkably, with only 50\% fully observed inputs and the rest partially observed, the recommendation performance of M\&Ms-VAE\textsubscript{$\mathbf{r}_u$} and M\&Ms-VAE\textsubscript{$\mathbf{r}_u,\mathbf{k}_u$} is decreased by only 9\% on average. More so, with 90\% partial observations, the model can still achieve more than 70\% of its performance quality on the full datasets. Finally, these results emphasize that M\&Ms-VAE can effectively learn the joint distribution even in a weakly supervised~setting. 

\section{Conclusion}
Recommendations can have much more impact if they are supported by explanations that can be critiqued.
Previous research has developed methods that either perform poorly in multi-step critiquing or suffer from computational inefficiency at inference.
In this paper, we presented M\&Ms-VAE, a novel variational autoencoder for recommendation~and explanation that treats the user preference and keyphrase usage as different observed variables. Additionally, we proposed a strategy that mimics weakly supervised learning and trains the inference networks jointly and~independently.

Our second contribution is a new critiquing module that leverages the generalizability of M\&Ms-VAE to embed the user preference and the critique. With a self-supervised objective and a synthetic dataset, we enable multi-step critiquing in M\&Ms-VAE.
Experiments on four datasets show that M\&Ms-VAE \begin{enumerate*}
 	\item is the first model to obtain  substantially better recommendation, explanation, and multi-critiquing performance,
 	\item processes critiques up to 25.6x faster than previous state-of-the-art methods, and
 	\item produces coherent joint and cross generation, even under weak~supervision.
 \end{enumerate*}
 \clearpage
 
\bibliographystyle{ACM-Reference-Format}

\clearpage
\appendix

\section{M\&Ms-VAE Derivation}

\begin{align}
\log~p(\bru, \bku) &= \log \int_{\bzu} p_\Theta(\bru, \bku, \bzu) d\bzu\\
&= \log \int_{\bzu}  p_\Theta(\bru, \bku, \bzu) \frac{q_\Phi(\bzu | \bru, \bku)}{q_\Phi(\bzu | \bru, \bku)} d\bzu\\
&= \log \biggl( \mathbb{E}_{q_\Phi(\bzu | \bru, \bku)} \biggl[ \frac{p_\Theta(\bru, \bku, \bzu)}{q_\Phi(\bzu | \bru, \bku)} \biggr] \biggr)\\
&\begin{aligned}
\geq \mathbb{E}_{q_\Phi(\bzu | \bru, \bku)} \bigl[\log p_\Theta(\bru, \bku, \bzu)\bigr] &- \mathbb{E}_{q_\Phi(\bzu | \bru, \bku)} \bigl[\log q_\Phi(\bzu | \bru, \bku) \bigr]
\end{aligned}\\
&\begin{aligned}
= \mathbb{E}_{q_\Phi(\bzu | \bru, \bku)} \bigl[\log p_\Theta(\bru, \bku | \bzu)\bigr] &+ \mathbb{E}_{q_\Phi(\bzu | \bru, \bku)} \bigl[\log p(\bzu) \bigr] - \mathbb{E}_{q_\Phi(\bzu | \bru, \bku)} \bigl[\log q_\Phi(\bzu | \bru, \bku) \bigr]\\
\end{aligned}\\
&\begin{aligned}
= \mathbb{E}_{q_\Phi(\bzu | \bru, \bku)} \bigl[\log p_\Theta(\bru, \bku | \bzu)\bigr] &- \textrm{D}_{\textrm{KL}} \bigl[ q_\Phi(\bzu | \bru, \bku) ~||~ p(\bzu) \bigr]\\
\end{aligned}\\
&\begin{aligned}
= \mathbb{E}_{q_\Phi(\bzu | \bru, \bku)} \bigl[\log p_{\Theta_r}(\bru | \bzu) + \log p_{\Theta_k}(\bku | \bzu)\bigr] &- \textrm{D}_{\textrm{KL}} \bigl[ q_\Phi(\bzu | \bru, \bku) ~||~ p(\bzu) \bigr]\\
\end{aligned}
\end{align}

\section{Keyphrase Examples}

\begin{table}[!h]
    \centering
   \caption{\label{tab_keyphrases}Some keyphrases mined from the reviews. We manually grouped them by types for a better understanding.}
\begin{threeparttable}
\begin{tabular}{@{}lllc@{}}
\textbf{Dataset} & \textbf{Type} & \textbf{Keyphrases}\\
\toprule
\multirow{4}{*}{Beer} & Head & white, tan, offwhite, brown\\
& Malt & roasted, caramel, pale, wheat, rye\\
& Color & golden, copper, orange, black, yellow\\
& Taste & citrus, fruit, chocolate, cherry, plum\\
\bottomrule
\multirow{4}{*}{CDs\&Vinyl} & Genre & rock, pop, jazz, rap, hip hop, R\&B\\
& Instrument & orchestra, drum\\
& Style & concert, opera\\
& Religious & chorus, christian, gospel\\
\bottomrule
\multirow{4}{*}{Yelp} & Cuisine & chinese, thai, italian, mexican, french\\
& Drink & tea, coffee, bubble tea, wine, soft drinks\\
& Food & chicken, beef, fish, pork, seafood, cheese\\
& Price \& Service & cheap, pricy, expensive, busy, friendly\\
\bottomrule
\multirow{4}{*}{Hotel} & Service & bar, lobby, housekeeping, guest, shuttle\\
& Cleanliness & toilet, sink, tub, smoking, toiletry, bathroom\\
& Location & airport, downtown, city, shop, restaurant\\
& Room & bed, tv, balcony, terrace, kitchen, business\\
\bottomrule
\end{tabular}
\end{threeparttable}
\end{table}

\section{Additional Training Details}

The official baselines' codes from the respective authors, including the tuning procedure, are available in \footnote{\url{https://github.com/wuga214/NCE_Projected_LRec}}\footnote{\url{https://github.com/k9luo/DeepCritiquingForVAEBasedRecSys}}\footnote{\url{https://github.com/wuga214/DeepCritiquingForRecSys}}\footnote{\url{https://github.com/litosly/RankingOptimizationApproachtoLLC}}. The~final hyperparameters for all models and datasets are shown in Table~\ref{table_parameters}. For all experiments, we used the following hardware: \begin{itemize*}
	\item \textbf{CPU}: 2x Intel Xeon E5-2680 v3 (Haswell), 2x 12 cores, 24 threads, 2.5 GHz, 30 MB cache;
	\item \textbf{RAM}: 16x16GB DDR4-2133;
	\item \textbf{GPU}: 1x Nvidia Titan X Maxwell;
	\item \textbf{OS}: Ubuntu 18.04;
	\item \textbf{Software}: Python 3.6, PyTorch 1.6.1, CUDA 10.2.
\end{itemize*}
\begin{table}[!h]
\small
    \centering
\caption{\label{table_parameters}Best hyperparameter setting for each model. The top table refers to Section~\ref{sec_rq1} and the bottom one to Section~\ref{sec_rq2}.}
\begin{threeparttable}
\begin{tabular}{@{}c@{\hspace{2mm}}lc@{\hspace{2mm}}c@{\hspace{2mm}}c@{\hspace{2mm}}c@{\hspace{2mm}}c@{\hspace{2mm}}c@{\hspace{2mm}}c@{\hspace{2mm}}c@{\hspace{2mm}}c@{\hspace{2mm}}c@{\hspace{2mm}}c@{\hspace{2mm}}c@{}}\\
\textbf{Dataset} & \textbf{Model} & $H$ & $LR$ & $\lambda_{L2}$ & $\lambda$ & $\lambda_{KP}$ & $\lambda_{C}$ & $\beta$ & Iteration & Epoch & Dropout & $\gamma$ & Neg. Samples\\
\toprule
\multirow{10}{*}{\rotatebox{90}{\textit{Beer}}}
& AutoRec & 200 & 0.0001 & 0.00001 & 1.0 & - & - &  - & - & 300 &-& - & -\\
& BPR & 200 & - & 0.0001 & 1.0 & - & - &  - & - & 30 &-& - & 1\\
& CDAE & 200 & 0.0001 & 0.00001 & 1.0 & - & - &  - & - & 300 & 0.2 & - & -\\
& NCE-PLRec & 50 & - & 10000.0 & 1.0 & - & - &  - & 10 & - &-& 1.1 & -\\
& PLRec & 400 & - & 10000.0 & 1.0 & - & - &  - & 10 & - &-& - & -\\
& PureSVD & 50 & - & - & - & - & - &  - & 10 & - &-& - & -\\
& VAE-CF & 50 & 0.0001 & 0.0001 & 1.0 & - & - &  0.2 & - & - & 0.4 & - & -\\
& CE-VAE & 100 & 0.0001 & 0.0001 & 1.0 & 0.01 & 0.01 &  0.001 & - & 300 & 0.5 & - & -\\
& CE-VNCF & 100 & 0.0005 & 0.00005 & 1.0 & 1.0 & 1.0 &  0.1 & - & 100 & 0.1 & - & 5\\
& M\&Ms-VAE & 300 & 0.00005 & 1e-10 & 3.0 & - & - & 0.7 & - & 300 & 0.4 & - & -\\
\bottomrule
\multirow{10}{*}{\rotatebox{90}{\textit{CDs\&Vinyl}}}
& AutoRec & 200 & 0.0001 & 0.00001 & 1.0 & - & - &  - & - & 300 &-& - & -\\
& BPR & 200 & - & 0.0001 & 1.0 & - & - &  - & - & 30 &-& - & 1\\
& CDAE & 200 & 0.0001 & 0.00001 & 1.0 & - & - &  - & - & 300 & 0.2 & - & -\\
& NCE-PLRec & 200 & - & 1000.0 & 1.0 & - & - &  - & 10 & - &-& 1.3 & -\\
& PLRec & 400 & - & 1000.0 & 1.0 & - & - &  - & 10 & - &-& - & -\\
& PureSVD & 200 & - & - & - & - &  - & 10 & - &-& - & -\\
& VAE-CF & 200 & 0.0001 & 0.00001 & 1.0 & - & - &  0.2 & - & - & 0.3 & - & -\\
& CE-VAE & 200 & 0.0001 & 0.0001 & 1.0 & 0.001 & 0.001 &  0.0001 & - & 600 & 0.5 & - & -\\
& CE-VNCF & 100 & 0.0001 & 0.0001 & 1.0 & 1.0 & 1.0 &  0.1 & - & 100 & 0.1 & - & 5\\
& M\&Ms-VAE & 400 & 0.00005 & 1e-12 & 1.0 & - & - & 0.4 & - & 400 & 0.4 & - & -\\
\bottomrule
\multirow{10}{*}{\rotatebox{90}{\textit{Yelp}}}
& AutoRec & 50 & 0.0001 & 0.001 & 1.0 & - & - &  - & - & 300 &-& - & -\\
& BPR & 100 & - & 0.0001 & 1.0 & - & - &  - & - & 30 &-& - & 1\\
& CDAE & 50 & 0.0001 & 0.001 & 1.0 & - & - &  - & - & 300 & 0.4 & - & -\\
& NCE-PLRec & 50 & - & 10000.0 & 1.0 & - & - &  - & 10 & - &-& 1.3 & -\\
& PLRec & 400 & - & 10000.0 & 1.0 & - & - &  - & 10 & - &-& - & -\\
& PureSVD & 50 & - & - & - & - & - &  - & 10 & - &-& - & -\\
& VAE-CF & 50 & 0.0001 & 0.001 & 1.0 & - & - &  0.2 & - & - & 0.2 & - & -\\
& CE-VAE & 200 & 0.0001 & 0.0001 & 1.0 & 0.01 & 0.01 &  0.001 & - & 600 & 0.4 & - & -\\
& CE-VNCF & 100 & 0.0005 & 0.0001 & 1.0 & 1.0 & 1.0 &  0.1 & - & 100 & 0.1 & - & 5\\
& M\&Ms-VAE & 500 & 0.00005 & 1e-10 & 10.0 & - & - & 0.8 & - & 300 & 0.7 & - & -\\
\bottomrule
\multirow{10}{*}{\rotatebox{90}{\textit{Hotel}}}
& AutoRec & 50 & 0.0001 & 1e-05 & 1.0 & - & - &  - & - & 300 &-& - & -\\
& BPR & 200 & - & 0.0001 & 1.0 & - & - &  - & - & 30 &-& - & 1\\
& CDAE & 200 & 0.0001 & 0.001 & 1.0 & - & - &  - & - & 300 & 0.2 & - & -\\
& NCE-PLRec & 50 & - & 10000.0 & 1.0 & - & - &  - & 10 & - &-& 1.3 & -\\
& PLRec & 400 & - & 10000.0 & 1.0 & - & - &  - & 10 & - &-& - & -\\
& PureSVD & 50 & - & - & - & - & - & - & 10 & - &-& - & -\\
& VAE-CF & 50 & 0.0001 & 1e-05 & 1.0 & - & - &  0.2 & - & - & 0.5 & - & -\\
& CE-VAE & 200 & 0.0001 & 0.0001 & 1.0 & 0.01 & 0.01 &  0.001 & - & 600 & 0.2 & - & -\\
& CE-VNCF & 100 & 0.0005 & 0.0001 & 1.0 & 1.0 & 1.0 &  0.1 & - & 100 & 0.1 & - & 5\\
& M\&Ms-VAE & 400 & 0.00005 & 1e-12 & 2.0 & - & - & 0.8 & - & 300 & - & - & -\\
\bottomrule
\end{tabular}
\end{threeparttable}
\begin{threeparttable}
\begin{tabular}{@{}lcccc@{}}\\
\textbf{Dataset} & \textbf{Model} & $h$ & $LR$ & $\lambda_{L2}$\\
\toprule
Beer & \multirow{4}{*}{\parbox{3cm}{\centering M\&Ms-VAE $\xi(\cdot)$\\(Critiquing)}} & 0.75 & 0.001 & 0\\
CDsVinyl &  & 3.0 & 0.001 & 1e-10\\
Yelp & & 2.0 & 0.001 & 0\\
Hotel &  & 5.0 & 0.001 & 1e-10\\
\bottomrule
\end{tabular}
\end{threeparttable}
\end{table}

\clearpage

\section{Multi-Step Critiquing on the whole set of items}

\begin{figure*}[!h]
\centering
\includegraphics[width=\textwidth,height=5.834in]{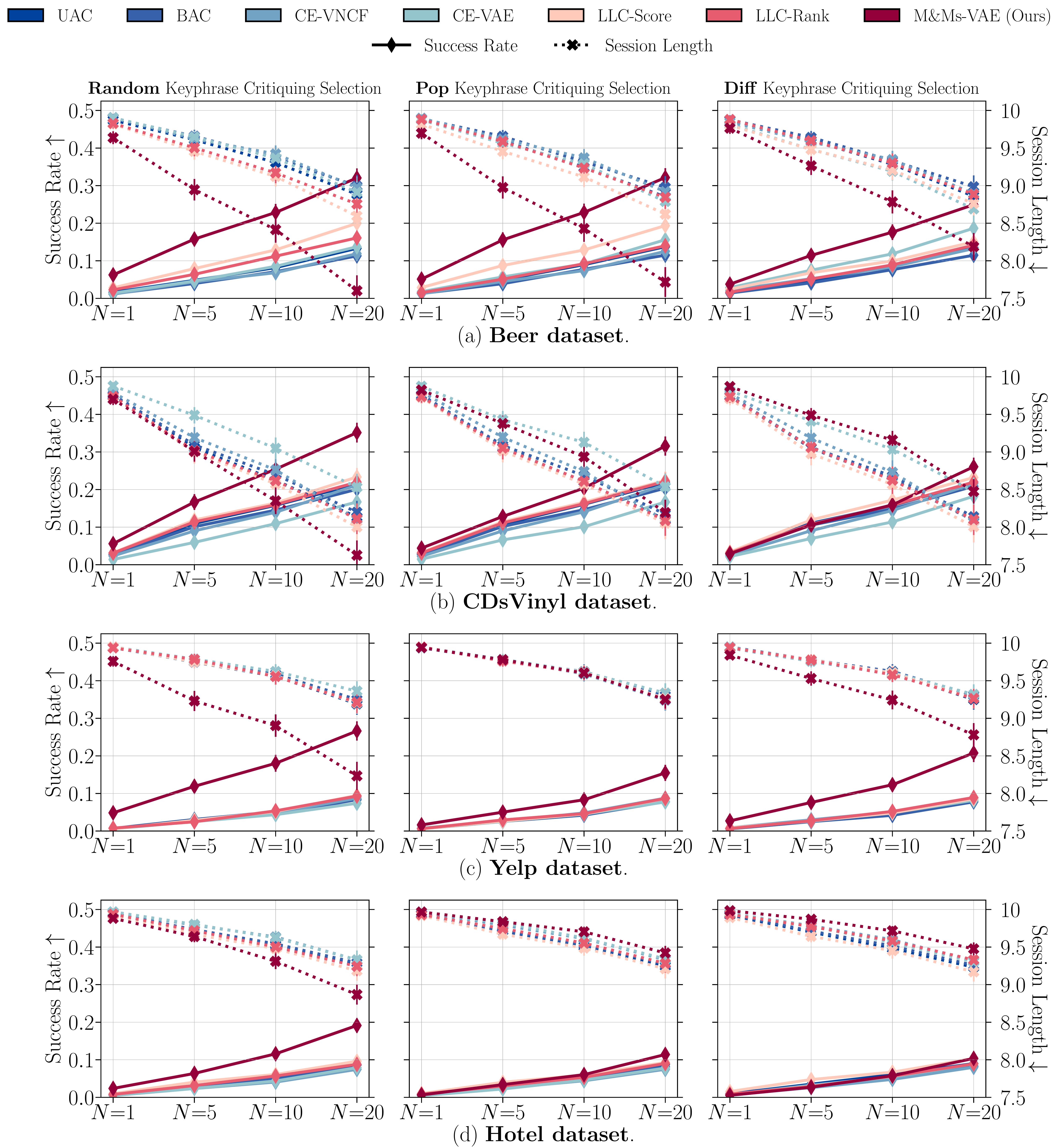}
\caption{\label{fig_multi_crit2}Multi-step critiquing performance after 10 turns of conversation on all items. For each dataset and keyphrase critiquing selection method, we report the average success rate (left y-axis) and session length (right y-axis) at different Top-N with 95\% confidence interval.}
\end{figure*}

Here we replicate the experiment in Section~\ref{sec_rq2}, but we use instead all the available items (see Table~\ref{stats_datasets} for the sizes). When the evaluation is conducted on 300 items (see Figure~\ref{fig_multi_crit}), we see that users indeed find a specific item using our technique with a high success rate (i.e., around 90\%). However, in Figure~\ref{fig_multi_crit2} where thousands of items are available,~the results show that current methods are not yet good enough to achieve similar results for such a large number. Nevertheless, M\&Ms-VAE clearly outperforms on average other methods and still achieves an average success rate~of~30\%. 

\end{document}